\documentclass[aps,pre,nobalancelastpage,superscriptaddress,twocolumn,longbibliography,nobibnotes]{revtex4-2}

\usepackage[utf8]{inputenc}
\usepackage[american,british]{babel}
\usepackage[T1]{fontenc}
\usepackage[pdftex]{graphicx}  
\usepackage[usenames, dvipsnames]{color}
\usepackage{xcolor}
\usepackage{dcolumn}
\usepackage{physics}
\usepackage{bm}
\usepackage{amsmath,amsthm,amssymb}
\usepackage{color}
\usepackage{verbatim} 
\usepackage[normalem]{ulem}

\usepackage{hyperref}
\hypersetup{
 colorlinks=true,
 linkcolor=blue,
 anchorcolor = blue,
 citecolor = blue,
 filecolor = blue,
 urlcolor = blue
}

\def \be {\begin{equation}} 
\def \ee {\end{equation}} 
\def \l {\left(} 
\def \r {\right)} 
\def \la {\langle} 
\def \ra {\rangle}

\newcommand{\lptms}{Universit\'e Paris-Saclay, CNRS, LPTMS, 91405, Orsay, France}
\newcommand{\osnabruck}{Department of Mathematics/Computer Science/Physics, University of Osnabr\"uck, D-49076 Osnabr\"uck, Germany} 
\newcommand{\smt}{Science, Mathematics and Technology Cluster, Singapore University of Technology and Design, 8 Somapah Road, 487372 Singapore}
\newcommand{\epd}{Engineering Product Development Pillar, Singapore University of Technology and Design, 8 Somapah Road, 487372 Singapore}
\newcommand{\cqt}{Centre for Quantum Technologies, National University of Singapore 117543, Singapore}
\newcommand{\majulab}{MajuLab, CNRS-UNS-NUS-NTU International Joint Research Unit, UMI 3654, Singapore}

\begin{document}

\title{Ergodicity and hydrodynamics: from quantum to classical spin systems}

\date{\today}

\author{Jiaozi Wang} 
\thanks{These authors contributed equally.} 
\affiliation{\osnabruck} 

\author{Luca Capizzi} 
\thanks{These authors contributed equally.}
\affiliation{\lptms}

\author{Dario Poletti}
\email{dario\_poletti@sutd.edu.sg}
\affiliation{\smt} 
\affiliation{\epd} 
\affiliation{\cqt} 
\affiliation{\majulab}

\author{Leonardo Mazza}
\email{leonardo.mazza@universite-paris-saclay.fr}
\affiliation{\lptms}

\begin{abstract}
We show that in classical spin systems the precise nature of the late-time hydrodynamic tails of the autocorrelation functions of a generic observable is determined by (i) the dynamical critical exponent and (ii) the equilibrium thermodynamic properties of the corresponding observable. 
We provide numerical results for one- and two-dimensional systems and present theoretical considerations that only rely on the notion of ergodicity.
Our result extends to the classical framework the \textit{relaxation-overlap inequality}, first introduced in [\textit{Capizzi et al.}~Phys.~Rev.~X~\textbf{15}, 011059 (2025)] for quantum many-body systems satisfying the eigenstate thermalization hypothesis.
\end{abstract}

\maketitle 

\section{Introduction}

Statistical mechanics is a cornerstone of our description of the physical world and is based on the crucial notion of \textit{ergodicity}, the fact that after a sufficiently long time, a physical system has explored the entire set of configurations that are allowed by its conservation laws, so that time averages correspond to phase space averages~\cite{Huang1987,Simon1993,Pathria2016}.
In macroscopic setups, ergodicity (see Refs.~\cite{Khinchin1949,ArnoldAvez1968,Ruelle1969}), can be employed to describe the long-time averages of relevant observables using statistical Gibbs ensembles, such as the canonical and grand canonical ones.

Thermodynamic ensembles do not evolve in time and are stationary. On the other hand, one is often interested in setting them slightly out of equilibrium with a perturbation and studying how this effect evolves in time and space. Specifically, the \textit{autocorrelation} functions of observables, and in particular their late-time properties, encode important information on such a dynamical behaviour. 
For instance, an exponential decay in time typically reflects a fast erasure of memory of an initially-applied perturbation~\cite{Petersen_1983,Ornstein_1989}, a phenomenology that is routinely observed in situations where no conservation laws are present.

In several common situations, the autocorrelation functions display a late-time algebraic decay to zero: this behaviour marks the appearance of an emergent \textit{hydrodynamic} behaviour that can be linked to the presence of conservation laws, such as the energy or the number of particles. Hydrodynamics allows to describe many-body systems in terms of local stationary states, where the temperature and the chemical potential varies smoothly in space and time: such a coarse-grained description turns out to be predictive at large scales, both for classical~\cite{km-63,ek-05,Spohn-12} and quantum~\cite{bkv-05,baw-06,nc-23,lmmr-14,mknsm-16,Crossno-16,lf-18,cdy-16,bcdf-16,dbd-19,Doyon-22,Doyon-22a} systems, by replacing microscopic complicated details with a few transport equations.

Focusing only on the simplest setups, those featuring energy as a unique conservation law, it is considered that the hydrodynamic tails should be only determined by the \textit{dynamical critical exponent} $z$, which is associated to the energy density spreading, and characterizes the universality class that is ballistic ($z=1$), diffusive ($z=2$), or else. 
This expectation is rooted in the fact that when energy is the only conserved quantity, its density spreads through the system, and the spreading of other generic local observables is typically determined by such a phenomenon; as such, the late-time dynamics of any observable is uniquely determined by the energy dynamics.

In a previous work~\cite{Capizzi_25} that we have co-authored (but see also Refs.~\cite{bbcp-21, bsrp-22, bp-22, DelacretazSciPost2020, wang2025eigenstate}), it was shown that for quantum many-body systems, the precise nature of the hydrodynamic tails of the autocorrelation functions of observables is linked to both (i) the dynamical critical exponent $z$ and (ii) the thermodynamic properties of the observable under analysis: such a relation was dubbed the \textit{relaxation-overlap inequality}. That derivation relied on the eigenstate thermalization hypothesis (ETH)~\cite{berry1977,Deutsch-91,Srednicki-99}, which formalizes the notion of chaotic quantum systems, and on its relation to hydrodynamics. 
While we expect that the main results in Ref.~\cite{Capizzi_25} should be valid in a larger framework including classical systems, many details that mostly rely on ETH need to be adapted to the classical framework.

A link between ETH and chaotic classical systems is well-known in the literature and it is provided by the well-established notion of \textit{ergodicity} (see Ref.~\cite{venuti2019ergodicity} and Refs.~\cite{mori_2017_ergodicity, Mori_Review_2018,  Alhambra_PRL2020}): roughly speaking, it says that the dynamics spreads uniformly along the accessible phase-space, giving rise to memory erasing phenomena and the late-time appearance of the microcanonical ensemble. 
We come back to this problem, giving a fresh view on the subject and we show, via the key assumption of ergodicity, how to extend the results of~\cite{Capizzi_25} to the classical framework.

In this article, we combine the aforementioned concepts and show with numerical and analytical tools that classical spin systems obey a relaxation-overlap inequality~\cite{Capizzi_25}.
The cornerstones of our derivation are (i) the notion of ergodicity and (ii) its relation to the hydrodynamic spreading of energy. 
Compared to quantum systems, predictions for classical spin systems can be studied numerically on larger sizes, corroborating the predictions with high-precision data.
We present those data focusing on several classical spin models characterized by different energy-spreading properties.
The results are fully compatible with a relaxation-overlap inequality, for which we propose an analytical derivation in a fully classical scenario.

The manuscript is organized as follows.
In Sec.~\ref{Sec:II} we outline the general mathematical framework that will be employed to study the classical spin systems, their thermodynamics, and their autocorrelation functions.
In Sec.~\ref{Sec:Numerics} we present numerical simulations for several spin models, defined in one-dimensional and two-dimensional lattices and in setups with different transport properties.
We show the existence of a link between the thermodynamical properties of certain observables and their autocorrelation functions.
In Sec.~\ref{Sec:Analytics}, we present some analytical considerations that prove such a connection.
Since similar results have already been identified in many-body \textit{quantum} spin systems, we present in Sec.~\ref{Sec:Discussion} a critical comparison of the two settings, focusing on the classical analogues of diagonal and off-diagonal ETH.
Finally, in Sec.~\ref{Sec:Conclusions}, we present our conclusions and a few perspectives of this work.

\section{Preliminaries}
\label{Sec:II}

In this section we provide some definitions useful for dealing with generic classical many-body systems, and specifically spin setups. The presentation follows a line that highlights similarities with the quantum setting.

\subsection{Generic classical systems}

We begin by describing classical systems in terms of observables, their dynamics, and introduce a notion of pure states and statistical mixtures.
Such a point of view, although slightly more abstract than elementary approaches, has the advantage of closely resembling the standard formulation of quantum systems, and therefore it allows us to bridge classical and quantum systems in a more direct way.

We start from a manifold $\mathcal{M}$ that we call \textit{phase-space} and we consider the \textit{observables} as real-valued smooth functions supported on $\mathcal{M}$, denoted by $C^{\infty}(\mathcal{M}, \mathbb R)$. In this context, a \textit{state} 
is a probability distribution over $\mathcal{M}$ and it allows to evaluate expectation values of observables. In particular, one can always associate \textit{pure states} to the points of $\mathcal{M}$: given a point $p\in \mathcal{M}$ one defines the state $\la \dots\ra_p$ as follows
\begin{equation}
\la \mathcal{O}\ra_p := \mathcal{O}(p), \quad \mathcal{O} \in C^{\infty}(\mathcal{M}, \mathbb R).
\end{equation}
These classical pure states do not have correlations since $\langle \mathcal O \mathcal O' \rangle_{p,c} = \la \mathcal{O}\mathcal{O}'\ra_{p} - \la \mathcal{O}\ra_p \la\mathcal{O}'\ra_p=0$ holds: this is a key difference with respect to quantum systems.
On the other hand, any statistical mixture of pure states is associated with a probability measure $dP_p$ on $\mathcal M$ which defines an expectation value
\begin{equation}
\la \dots \ra := \int_{\mathcal{M}} dP_p \la \dots\ra_p.
\end{equation}
In particular, equilibrium states associated with the microcanonical, canonical or grandcanonical ensembles are statistical mixtures; as it is well known, they
show correlations whose origin is purely classical.

The \textit{dynamics of classical} system can be described in terms of the evolution of observables. For simplicity, here we only consider dynamics that arise from a Poisson-bracket structure and a Hamiltonian $H$ which, for the sake of simplicity, does not depend on time. The evolution of an observable $\mathcal{O}$ is described by the equation
\begin{equation}\label{eq:O_evol}
\frac{d}{dt}\mathcal{O} = \{H,\mathcal{O}\},
\end{equation}
which allows us to associate to any observable $\mathcal{O}$ its time-evolved counterpart $\mathcal{O}(t)$. As it is obvious, in a time-independent system energy is conserved, namely $\frac d {dt} H = 0$ because the Poisson brackets are required to be antisymmetric in the two variables. Also, a \textit{stationary state} $\la \dots\ra$ satisfies, by definition, $ \langle \mathcal{O}(t)\rangle = \langle \mathcal O \rangle, \forall \mathcal{O}$, or, which is equivalent, $ \frac{d}{dt}\la \mathcal{O}(t)\ra =0, \forall \mathcal{O}$. This viewpoint is analogous to the \textit{Heisenberg picture} in quantum mechanics. 

\subsection{Spin systems}

In this work, we will focus on classical spin systems. Their interest comes from the fact that they correctly describe the large-$S$ limit of quantum spin-$S$ systems; 
in general, they are expected to capture the same relevant physics (for instance, diffusive hydrodynamics of energy) of quantum spin systems when $S$ is finite (see for instance the intriguing results on the comparison of autocorrelation functions of classical and quantum spin setups in Ref.~\cite{Schubert-21}): this is important because it motivates the attempt presented in this article to relate the findings of~\cite{Capizzi_25}, tested on a spin-$1$ quantum Ising chain, to possible classical counterparts.
 
We first discuss the phase space of a single spin. We identify the phase space as $\mathcal{M} := \mathbb{R}^3$, where the $3$ coordindates are the three directions of the spin, $S^x$, $S^y$ and $S^z$. One introduces the Poisson brackets on $\mathcal{M}$ satisfying for instance $\{S^x,S^y\} = S^z$ and more generically
\begin{equation}
\{S^\mu,S^\nu\} = \epsilon_{\mu \nu \lambda}S^\lambda,
\end{equation}
where $\epsilon_{\mu \nu \lambda}$ is the fully-antisymmetric Levi-Civita symbol.
One can show that the squared modulus $(S^x)^2+(S^y)^2+(S^z)^2$, known as the \textit{Casimir invariant}, is conserved under time evolution no matter the specific choice of the Hamiltonian: for this reason, the dynamics is restricted on submanifolds at fixed values of the Casimir invariant.

We can associate a rotational invariant measure (known as the \textit{Haar measure}) to the spherical surface $(S^x)^2+(S^y)^2+(S^z)^2=1$, whose associated state plays the role of an infinite temperature state, i.e. $\beta=0$: this is the only state with the property $\la \{S^a, \mathcal{O}\}\ra_{\beta=0} = 0$ for $a=x,y,z$ and for any observable $\mathcal O$.
Hence, the infinite-temperature state is stationary.
With a bit of algebra, one can characterize systematically the expectation values of observables in the infinite temperature states and obtain, for example, that $  \la (S^a)^2\ra_{\beta = 0} = 1/3$. Further details are given in Appendix \ref{app:exp_val}.

These building blocks allow to construct in a natural way the phase space and infinite temperature state of a system with many classical spins. Important properties for the following are that the infinite temperature state is stationary and it does not correlate distinct spins, namely $\la \mathcal{O}(\mathbf r)\mathcal{O}'(\mathbf r')\ra_{\beta=0,c}=0$ for $\mathbf {r} \neq \mathbf {r}'$: the same properties hold for quantum mechanical systems as well.

\subsection{Thermodynamics}

In the previous paragraph we have introduced the infinite temperature state $\la \dots\ra_{\beta=0}$, which can be considered as a universal reference stationary measure on the phase space.
Similarly, one can construct a \textit{thermal state} associated with a Hamiltonian $H$ as
\begin{equation}
\la \dots \ra_\beta := \frac{\la \dots e^{-\beta H}\ra_{\beta'=0}}{\la e^{-\beta H}\ra_{\beta'=0}}.
\end{equation}
One can easily prove that the thermal state is stationary for the dynamics induced by $H$: this comes directly from the stationarity of the infinite temperature state and that of $H$ ($H(t) = H$).

We can hence define the thermal expectation of an observable $\mathcal O$ as:
\begin{equation}
    \mathcal O (\beta) := \langle \mathcal O \rangle_\beta
\end{equation}
General mathematical results on spin systems \cite{fv-17}, show that for infinite lattices with short-range interactions, the thermal states (at sufficiently high temperature, above any possible low-temperature symmetry breaking) satisfy the \textit{clustering property}: the connected correlation functions $\la \mathcal{O}(\mathbf r)\mathcal{O}'(\mathbf r')\ra_{\beta,c}$ decay exponentially fast in the distance $|\mathbf r - \mathbf r'|$.
Also, as a consequence, the cumulants of extensive observables, namely those that can be written as sums $\mathcal Q= \sum_{\bf r} q(\mathbf r)$, with $q(\mathbf r)$ the corresponding charge density, are extensive:
for a system of volume $V$, then $\la \mathcal Q^n\ra_{\beta,c} \sim V$ is true and the central limit theorem holds (here $n$ is a positive integer and $V$ is the number of spins).    

Since for many-body systems the Hamiltonian is an extensive observable, it is convenient to define the energy density of a thermal state as
\begin{equation}
    \varepsilon (\beta) := \frac{1}{V} \langle H \rangle_\beta.
\end{equation}
We expect $\varepsilon(\beta)$ to be invertible: technically, this is guaranteed whenever the entropy is a strictly-convex function of the energy and the heat capacity is positive. As a consequence, one can equivalently parameterize the thermal state either with $\beta$ or with the corresponding energy density $\varepsilon(\beta)$. In particular, the expectation value of any observable $\mathcal O$ at a given energy density $\varepsilon$
\begin{equation}
    \mathcal O(\varepsilon) := \langle \mathcal O \rangle_{\beta(\varepsilon)}
    \label{Eq:EnergyDependence}
\end{equation}
will play an important role in our discussion.

Lastly, for completeness, we recall that, as long as expectation values of local observables are implied, the microcanonical and canonical ensembles become equivalent in the thermodynamic limit; specifically, $\mathcal O(\varepsilon)$ gives the microcanonical expectation value at a given energy density. Similarly, such an object enters the diagonal matrix elements of the ETH for quantum chaotic systems.

\subsection{Autocorrelation functions}

A common way of characterizing the dynamical properties of an equilibrium state is to compute multipoint correlation functions of local observables. In this work we focus on the autocorrelation function of thermal states, defined as
\begin{equation}
C(t) = 
\la \mathcal{O}(t)\mathcal{O}\ra_{\beta,c}:=
\la \mathcal{O}(t)\mathcal{O}\ra_{\beta}
-
\la \mathcal{O}(t) \rangle_\beta \langle\mathcal{O}\ra_{\beta}.
\label{Eq:Autocorrelation}
\end{equation}
The late-time behaviour of such a quantity encodes relevant features of the underlying energy hydrodynamics, and it is associated with hydrodynamic tails of the form $\la \mathcal{O}(t)\mathcal{O}\ra_{\beta,c} \sim t^{-\nu}$; this scaling is expected to occur at large time in the infinite volume limit.

The exponent $\nu$ is an emergent feature of the dynamics of the operator $\mathcal{O}$. For example, in the framework of this article, where the energy is the only conserved charge, a relevant operator is the energy density $h$, defined by $H = \sum_{\bf r} h(\mathbf r)$; in that case,  $\nu = d/z$, where $z$ is the \textit{dynamical critical exponent} \cite{Spohn-12} of the energy transport and $d$ the dimensionality of the setup. Common values for local hamiltonians are $z=1$, corresponding to situations where energy spreads ballistically, and $z=2$, where it spreads diffusively.
The former is typically associated with integrable models, and hence will not be encountered here, while the latter occurs in generic (chaotic) local systems.
Furthermore, as we will see, when considering long-range interactions, other values of $z$ can appear.

\section{Numerical analysis}
\label{Sec:Numerics}

We begin our discussion by presenting some numerical results for one- and two-dimensional classical spin systems.
Our goal is to highlight a connection between the hydrodynamic tail of the autocorrelation function of an observable $\mathcal O$ and its energy dependence as expressed by Eq.~\eqref{Eq:EnergyDependence}.
Our numerical results will focus on the infinite-temperature state $\beta=0$.


\subsection{The models}

The first model that we consider is a one-dimensional Ising chain with tilted field,
\begin{equation}
H=\sum_{j}h_{x}S_{j}^{x}+h_{z}S_{j}^{z}+JS_{j}^{z}S_{j+1}^{z},
\label{Eq:H:Ising}
\end{equation}
where $S^{a}_j$ denotes the components $a=x,y,z$ of a classical spin at site $j$.
The chosen parameters are $h_x = 1.1,\ h_z = 0.9$ and $J = 1$. 

The second model is a one-dimensional Ising chain with transverse field and long-range couplings,
\begin{subequations}
\label{Eq:H:Ising:LR}
\begin{equation}
H=\frac{1}{2}\sum_{i}\sum_{j\neq i}\frac{J}{N_{\alpha}r_{ij}^{\alpha}}S_{i}^{z}S_{j}^{z}+\sum_{i}h_{x}S_{i}^{x},
\end{equation}
where
\begin{equation}
    r_{ij}=\min(|j-i|,L-|j-i|),\ N_{\alpha}=\left(\sum_{i=2}^{L}\frac{1}{r_{i1}^{2\alpha}}\right)^{-\frac 12}.
\end{equation}
\end{subequations}
The energy transport is diffusive for $\alpha \ge 1.5$
and becomes anomalous at $1 <\alpha < 1.5$ with a dynamical exponent $z = 2\alpha - 1$ \cite{nishikawa2025energy-LR-Saito25}. Here we choose $h_x = 1.1,\ J = 2\sqrt{2}$ and two different values of $\alpha$: $\alpha = 1.1$ and $\alpha = 1.5$.

Finally, as a third model, we consider a two-dimensional Ising model on a  $\ell \times \ell$ square lattice,
\begin{equation}
    H=\sum_{\boldsymbol{i}}h_{x}S_{\boldsymbol{i}}^{x}+\sum_{\langle \boldsymbol{ij}\rangle}JS_{\boldsymbol{i}}^{z}S_{\boldsymbol{j}}^{z};
    \label{Eq:H:Ising:2D}
\end{equation}
where $\boldsymbol{i}$ and $\boldsymbol{j}$ label lattice sites and the sum over $\langle \boldsymbol{ij} \rangle$ runs over bonds connecting neighboring sites and we choose $h_x = 1.1$ and $J = 1$. 
In all setups we consider periodic boundary conditions. 


\subsection{Numerical technique}
In the calculation of thermal expectation value $\langle {\cal O} \rangle_\beta$, we employ a Hamiltonian Monte-Carlo methods with a  Metropolis-Hastings algorithm \cite{metropolis1953equation,hastings1970monte}.
The thermal average is taken over 
$10^8$ different configurations for each given temperature $T$.

In the calculation of auto-correlation function $C(t)$,  
we employ a Yoshida fourth-order symplectic integrator \cite{Yoshida}
with a time step of $\delta t = 0.02$ ($\delta t = 0.01$ for tilted field Ising model). The initial spin configurations
are sampled using the Monte Carlo method (at temperature $T$ where only infinite temperature $T = \infty$ is considered here). The results are averaged over $10^7$ independent initial configurations.


\subsection{Numerical results}

\begin{figure}[t]
	\centering
    \includegraphics[width = 1\linewidth]{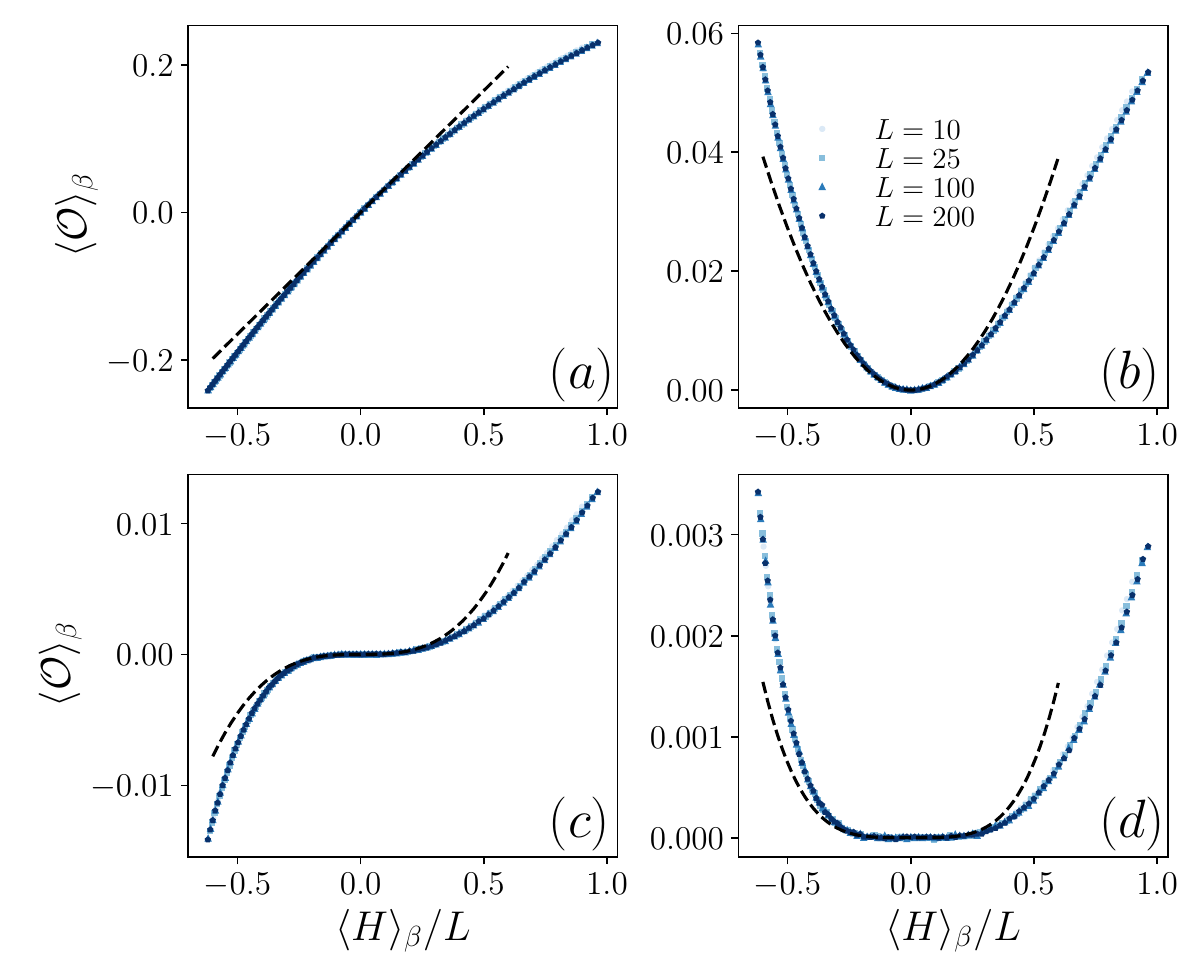}
	\caption{$\langle {\cal O} \rangle_\beta$ versus $\langle {H} \rangle_\beta/L$  for the observables (a) $S^x_j$; (b) $S^x_j S^x_{j+1}$; (c) $S^x_j S^x_{j+1}S^x_{j+2}$ and (d) $S^x_j S^x_{j +1} S^x_{j+2}S^x_{j+3}$ in the one-dimensional Ising model with tilted field. The dashed line indicates the analytical prediction $\varepsilon^m$ derived in Appendix~\ref{app:Overlaps}.
	}\label{Fig:Ising:Thermal}

    \includegraphics[width =\columnwidth]{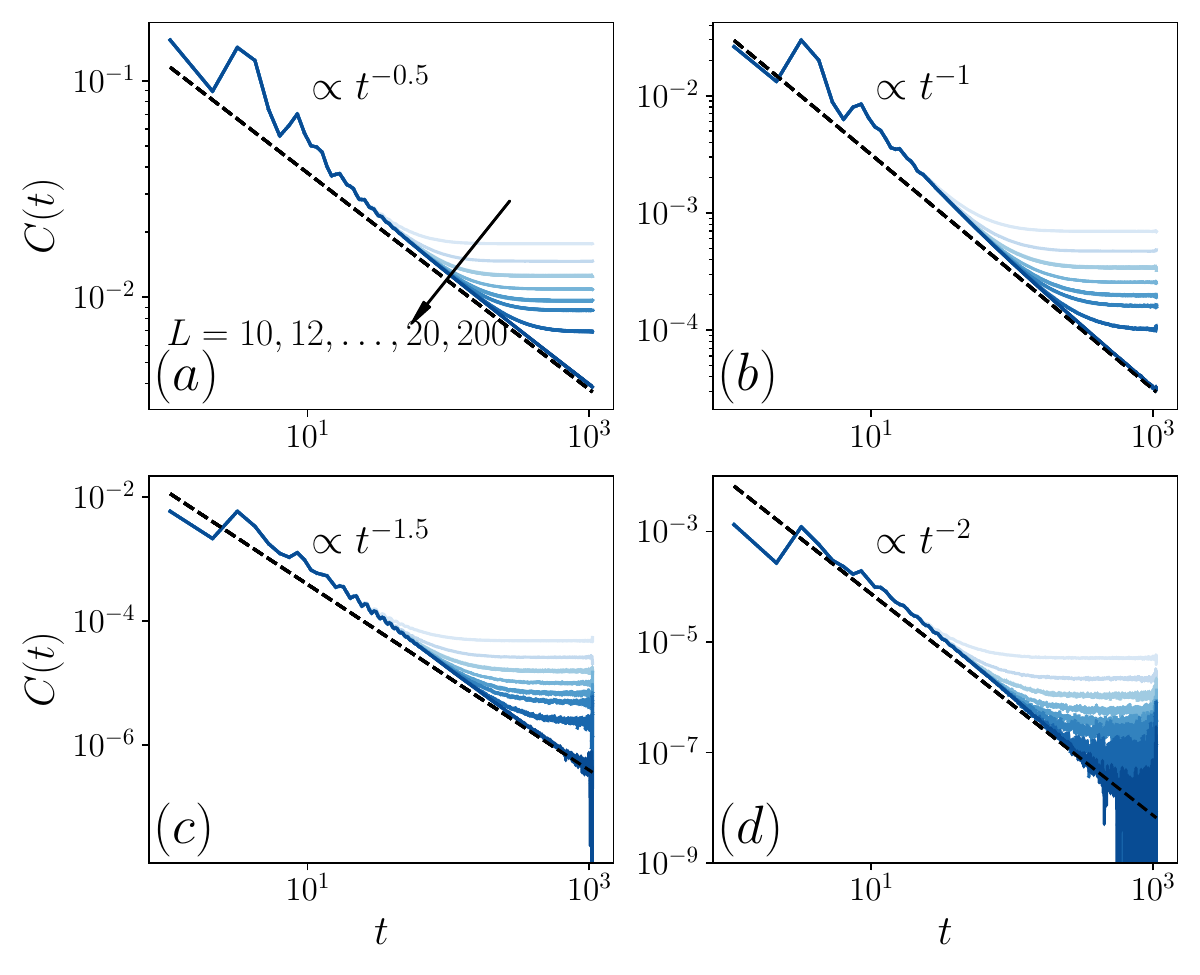}
	\caption{$C(t)$ versus $t$ for the infinite temperature $\beta = 0$ state for the four observables (a) $S^x_j$; (b) $S^x_j S^x_{j+1}$; (c) $S^x_j S^x_{j+1}S^x_{j+2}$ and (d) $S^x_j S^x_{j +1} S^x_{j+2}S^x_{j+3}$ in the one-dimensional  Ising model with tilted field. Dashed line indicate $\propto t^{dm/z}$ for $m=1,2,3,4$, respectively with $z=2$ and $d= 1$.}
	\label{Fig:Ising:AutoC}
\end{figure}

We start by examining the thermodynamic properties of four observables, $S_j^x$, $S_j^x S_{j+1}^x$,
$S_j^x S_{j+1}^x S_{j+2}^x$ and
$S_j^x S_{j+1}^x S_{j+2}^x S_{j+3}^x$ for the Ising model with tilted field in Eq.~\eqref{Eq:H:Ising}.
In Fig.~\ref{Fig:Ising:Thermal} we show the energy dependence of the four observables, $\mathcal O(\varepsilon)$, computed according to the definition in Eq.~\eqref{Eq:EnergyDependence}.
We consider a system of size up to $L = 200$, which is sufficient to highlight the infinite-size behaviour and make $1/L$ corrections negligible on the scales of the plot.
The observables have been chosen because the functions $\mathcal O(\varepsilon)$ display markedly different behaviours around $\varepsilon =0$, and
specifically, a small-energy behaviour $\sim \varepsilon^m$ with $m=1$, $2$, $3$ and $4$, respectively; the four scalings are proven analytically in Appendix \ref{app:Overlaps}.

\begin{figure}[t]
	\centering
    \includegraphics[width = 1\linewidth]{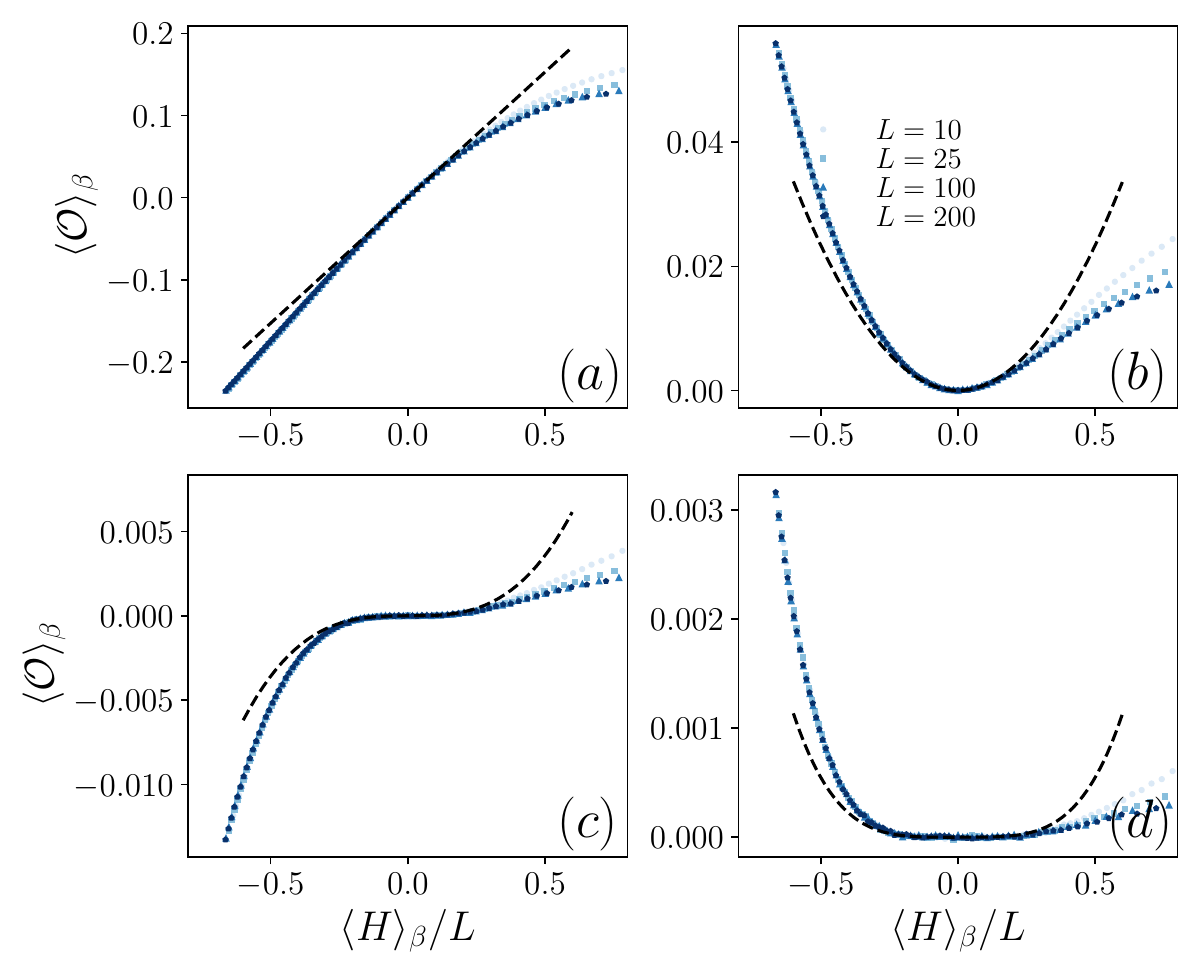}
	\caption{$\langle {\cal O} \rangle_\beta$ versus $\langle {H} \rangle_\beta/L$  for observables (a) $S^x_j$; (b) $S^x_j S^x_{j+1}$; (c) $S^x_j S^x_{j+1}S^x_{j+2}$ and (d) $S^x_j S^x_{j +1} S^x_{j+2}S^x_{j+3}$ in the one-dimensional long range ($\alpha = 1.5$) Ising model with tilted field. The dashed line indicates the analytical prediction $\varepsilon^m$ derived in Appendix~\ref{app:Overlaps}.
	}\label{Fig:Ising-LR:Thermal}

    \includegraphics[width =\columnwidth]{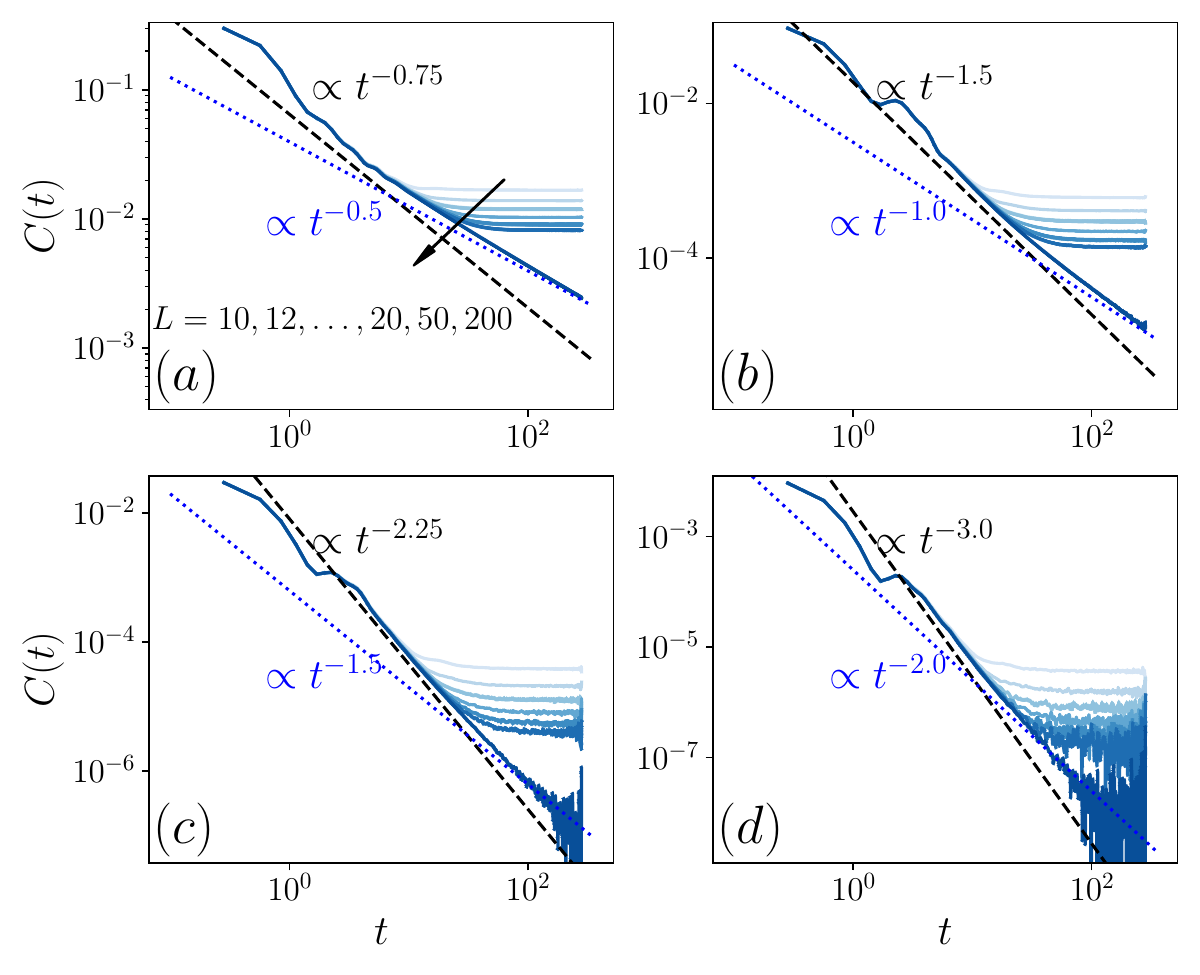}
	\caption{$C(t)$ versus $t$ for infinite temperature $\beta = 0$ for observables (a) $S^x_j$; (b) $S^x_j S^x_{j+1}$; (c) $S^x_j S^x_{j+1}S^x_{j+2}$ and (d) $S^x_j S^x_{j +1} S^x_{j+2}S^x_{j+3}$ in one-dimensional long-range Ising model ($\alpha = 1.5$). The black dashed line indicate $\propto t^{dm/z}$ for $m=1,2,3,4$, respectively with $z=4/3$ (used in Ref.~\cite{Capizzi_25}) and $d= 1$.
    The blue dotted line indicate $\propto t^{dm/z}$ with $z=2$ and $d= 1$ for comparison (prediction of Ref.~\cite{nishikawa2025energy-LR-Saito25}).
    }
	\label{Fig:Ising-LR:AutoC}
\end{figure}

\begin{figure}[t]
	\centering
    \includegraphics[width = 1\linewidth]{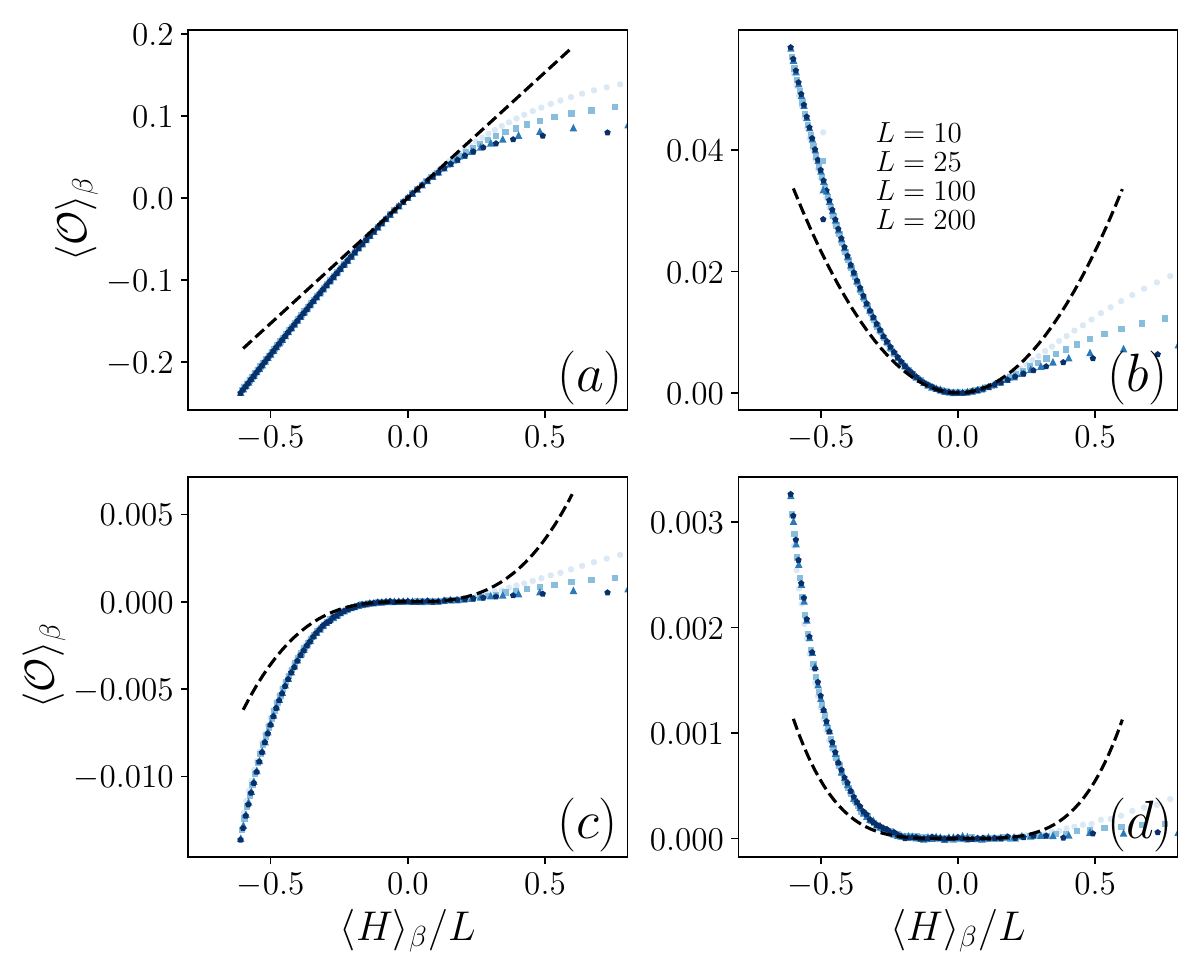}
	\caption{$\langle {\cal O} \rangle_\beta$ versus $\langle {H} \rangle_\beta/L$  for observables (a) $S^x_j$; (b) $S^x_j S^x_{j+1}$; (c) $S^x_j S^x_{j+1}S^x_{j+2}$ and (d) $S^x_j S^x_{j +1} S^x_{j+2}S^x_{j+3}$ in the one-dimensional long range ($\alpha = 1.1$) Ising model with tilted field. The dashed line indicates the analytical prediction $\varepsilon^m$ derived in Appendix~\ref{app:Overlaps}.
	}\label{Fig:Ising-LR:Thermal-1.1}

    \includegraphics[width =\columnwidth]{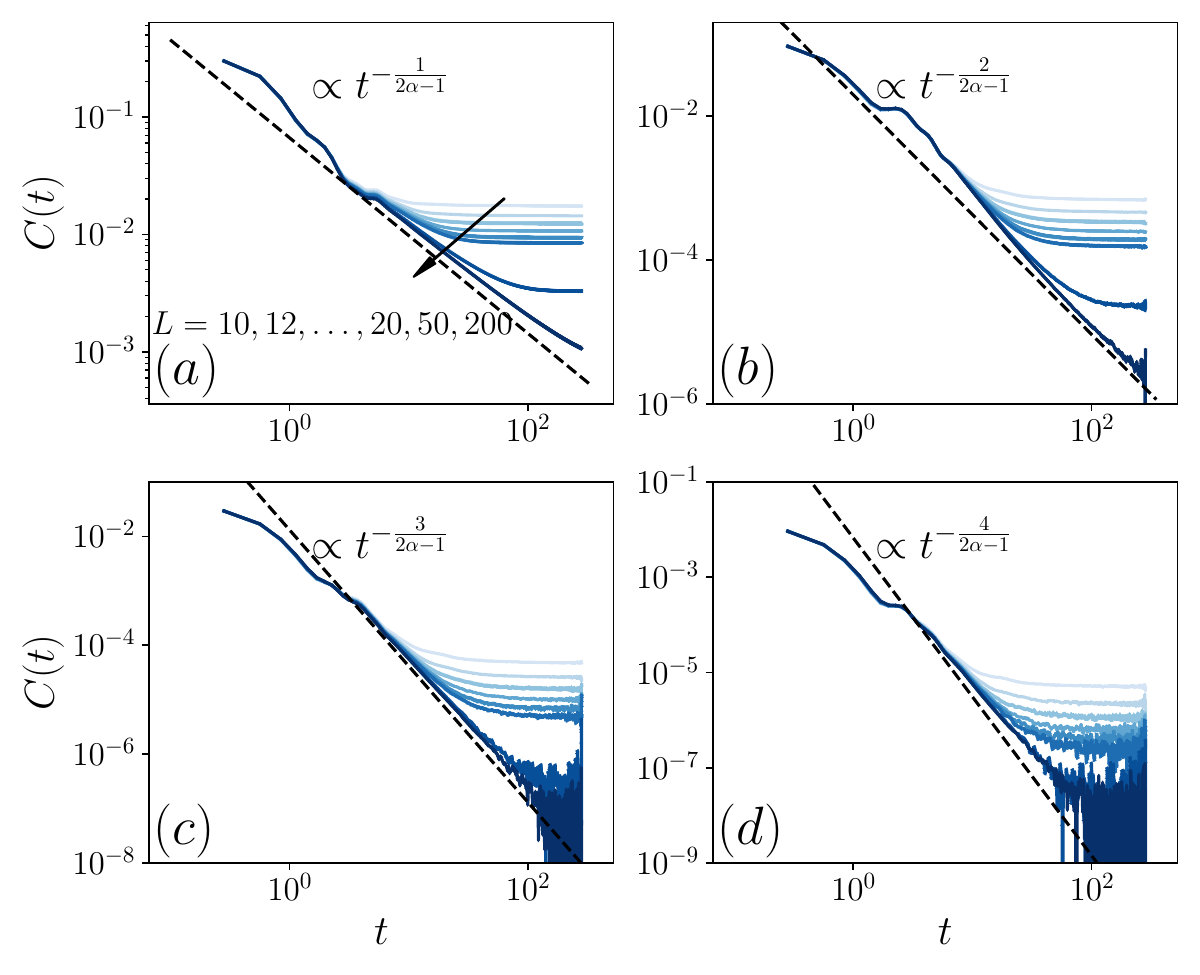}
	\caption{$C(t)$ versus $t$ for infinite temperature $\beta = 0$ for observables (a) $S^x_j$; (b) $S^x_j S^x_{j+1}$; (c) $S^x_j S^x_{j+1}S^x_{j+2}$ and (d) $S^x_j S^x_{j +1} S^x_{j+2}S^x_{j+3}$ in one-dimensional long-range Ising model ($\alpha = 1.1$). The black dashed line indicate indicate $\propto t^{dm/z}$ with $z=(2\alpha -1)$ and $d = 1$ (prediction of Ref.~\cite{nishikawa2025energy-LR-Saito25}).
    }
	\label{Fig:Ising-LR:AutoC-1.1}
\end{figure}

The behavior of $\mathcal O(\varepsilon)$ at energy density $\varepsilon=0$ is particularly relevant because we will now numerically investigate the dynamical properties of the infinite-temperature state, which has energy density $\varepsilon = 0$ (this is true in all three models);
specifically, we compute the autocorrelation function $C(t)$ defined in Eq.~\eqref{Eq:Autocorrelation} of the same four observables on the thermal state with $\beta=0$.
In the four panels of Fig.~\ref{Fig:Ising:AutoC} we show the results of the calculations performed for several spin-chain lengths, from $L=10$ to $L=200$.
In all cases, three time-regimes can be identified.
First, we observe an initial transient with oscillations, which extends up to time-scales of order $\tau_1 \sim O(1)$.
Afterwards we recognize the onset of a clear algebraic decay, interpreted as a hydrodynamic tail.
Finally, the last time-regime is characterized by a stationary behavior, and the autocorrelation function does not display a significant dependence on time.
This latter regime is clearly a finite-size effect and indeed starts at a time $\tau_2$ that scales with the chain length $L$, since the plateaus depart from the same universal power-law decay at times that increase the system size.
As a consequence, the plateau values depend on $L$.

The exponent $\nu$ of the hydrodynamic tail depends on the observable. The values that we have identified are compatible with $1/2$, $1$, $3/2$ and $2$ for the four observables, respectively, and are highlighted by a black dashed line in Fig.~\ref{Fig:Ising:AutoC};
the theory that we will present in Sec.~\ref{Sec:Analytics} will propose exactly these fractions.
At this stage, it is possible to guess a link with the value $m$ that characterizes the thermodynamic function $\mathcal O(\varepsilon)$, namely $\nu=m/2$.

A similar analysis is presented in Figs.~\ref{Fig:Ising-LR:Thermal} and~\ref{Fig:Ising-LR:AutoC} for the one-dimensional long-range Ising model of Eq.~\eqref{Eq:H:Ising:LR} with long-range coupling set by the parameter $\alpha = 1.5$.
By considering the same four observables, we show the behavior of $\mathcal O(\varepsilon) \sim \varepsilon^m$ with $m$ varying from $1$ to $4$ as in the short-range model (see also Appendix~\ref{app:Overlaps}).
The autocorrelation functions are more interesting and display two different algebraic decays (see also Ref.~\cite{nishikawa2025energy-LR-Saito25} for an in-depth study of the autocorrelation functions of this model).
A first hydrodynamic tail $t^{- \nu^{(1)}}$ highlighted by a black dashed line is compatible with the values $\nu^{(1)} = 0.75$, $1.5$, $2.25$ and $3$ and can be summarized by the formula $\nu^{(1)} = 3 m / 4 $; this behavior is well highlighted by our numerics for setups with lengths up to $L=20$ and it coincides exactly with the result that we co-presented in Ref.~\cite{Capizzi_25} for the corresponding quantum spin-1 setup.
By exploiting the possibility offered by classical systems of studying larger setups for longer times, we investigate a system of length $L=200$ and discover that the power-law bends at late times towards a novel hydrodynamic tail $t^{-\nu^{(2)}}$ with $\nu^{(2)}=0.5$, $1$, $1.5$ and $2$, highlighted in the figure by a blue dotted line.
This late-time result is predicted in Ref.~\cite{nishikawa2025energy-LR-Saito25}.

By changing the range of the interactions setting $\alpha = 1.1$, the analysis presented in Figs.~\ref{Fig:Ising-LR:Thermal-1.1}
and~\ref{Fig:Ising-LR:AutoC-1.1}
shows that as long as the thermal expectation values of the four observables are considered, the behaviours are fully identical to those encountered so far.
The autocorrelation functions are instead displaying one algebraic hydrodynamic tail, that is compatible with the value $\nu = m /(2 \alpha-1)$ highlighted by the black dashed line. 
\begin{figure}[t]
	\centering
    \includegraphics[width = 1\linewidth]{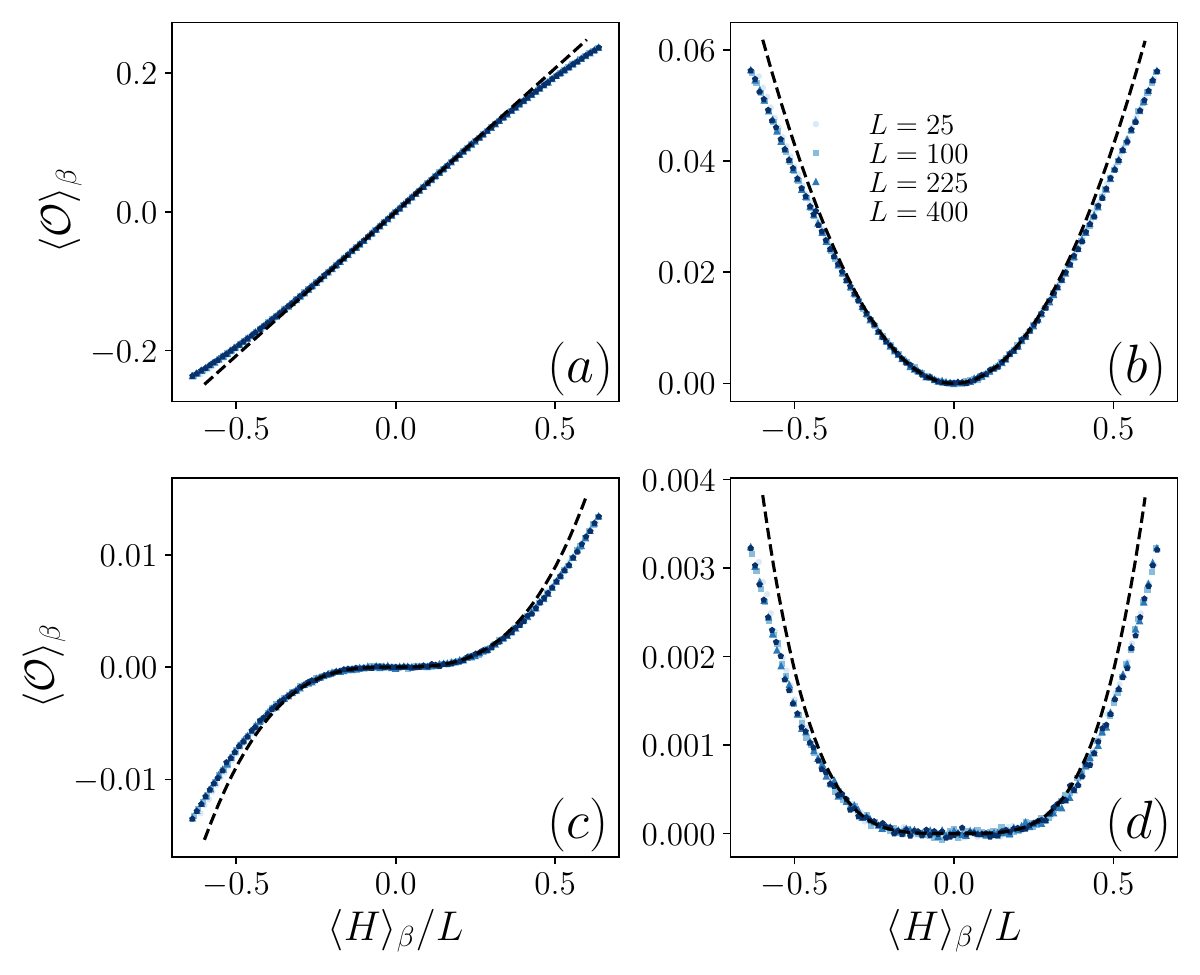}
	\caption{$\langle {\cal O} \rangle_\beta$ versus $\langle {H} \rangle_\beta/L$  for observables (a) $S^x_j$; (b) $S^x_j S^x_{j+1}$; (c) $S^x_j S^x_{j+1}S^x_{j+2}$ and (d) $S^x_j S^x_{j +1} S^x_{j+2}S^x_{j+3}$ in the two-dimensionael Ising model. The dashed line indicates the analytical prediction $\varepsilon^m$ derived in Appendix~\ref{app:Overlaps}.
	}\label{Fig:Ising-2D:Thermal}

    \includegraphics[width =\columnwidth]{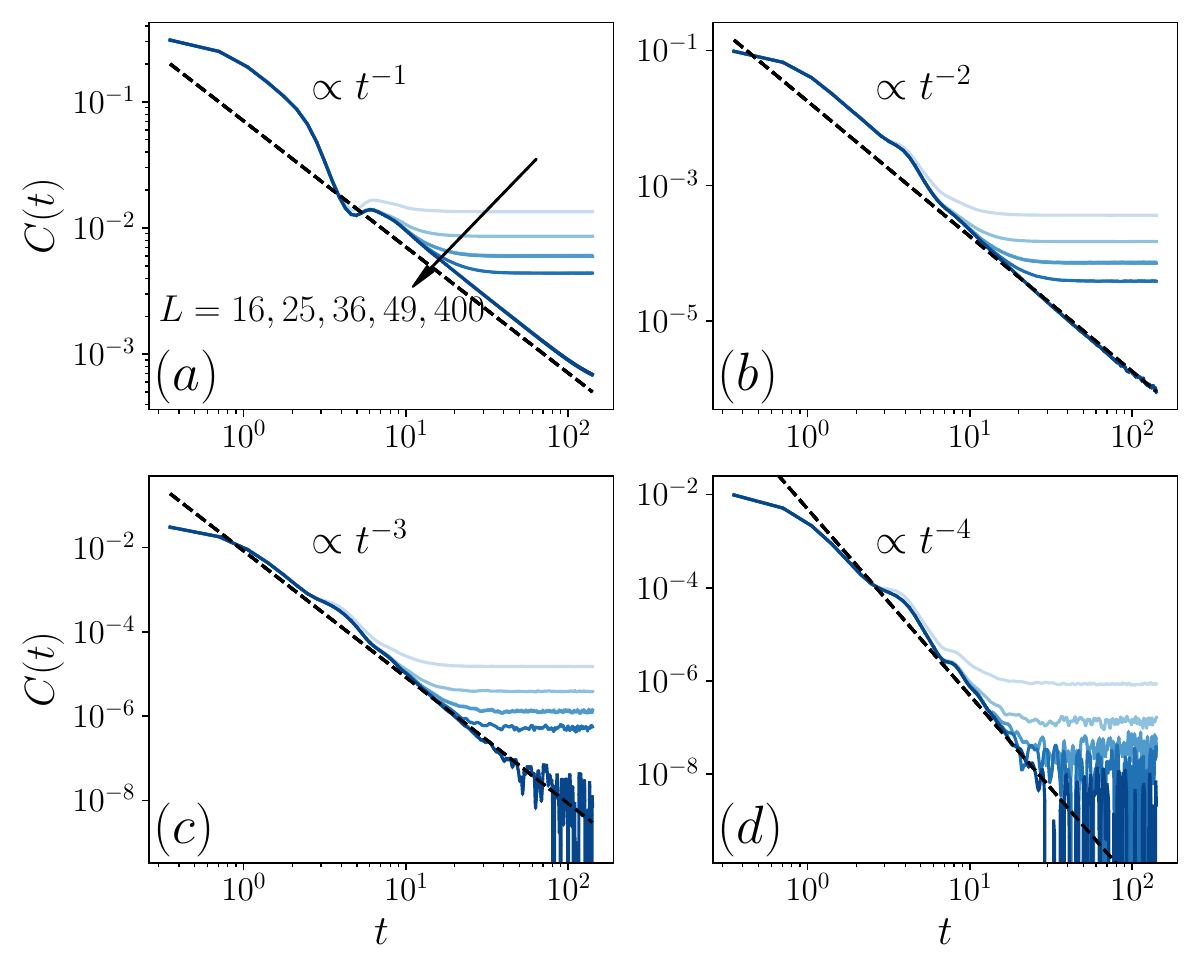}
	\caption{$C(t)$ versus $t$ for infinite temperature $\beta = 0$ for observables (a) $S^x_j$; (b) $S^x_j S^x_{j+1}$; (c) $S^x_j S^x_{j+1}S^x_{j+2}$ and (d) $S^x_j S^x_{j +1} S^x_{j+2}S^x_{j+3}$ in the two-dimensional transverse Ising model ($\ell \times \ell$ square lattice). Here $L = \ell^2$, which indicates the total number of sites. Dashed line indicate $\propto t^{dm/2}$ for $m=1,2,3,4$, respectively with $z=2$ and $d = 2$.}
	\label{Fig:Ising-2D:AutoC}
\end{figure}

Finally, in Figs.~\ref{Fig:Ising-2D:Thermal} and~\ref{Fig:Ising-2D:AutoC}
we present the same analysis for the two-dimensional Ising model in Eq.~\eqref{Eq:H:Ising:2D}.
Whereas the thermodynamic properties of the model, represented by the function $\mathcal O(\varepsilon)$, are essentially coinciding with those of the two previous models, the autocorrelation functions display a marked intermediate-time hydrodynamic tail, $t^{-\nu}$. For this model, the values that we observe are compatible with $\nu=1$, $2$, $3$ and $4$ and can be summarized by the relation $\nu = m$.

To conclude, 
two interesting quantitative properties of the autocorrelation functions emerge from this numerical analysis and must be stressed.
First, the exponents $\nu$ that we have fitted are all compatible with the following formula 
\begin{equation}
\nu = \frac{d m} {z}, 
\label{Eq:Ov:Rel}
\end{equation}
where $d$ is the dimensionality, $m$ reflects the thermodynamic behavior $\mathcal O(\varepsilon) \sim \varepsilon^m$ and $z$ will be identified in Sec.~\ref{Sec:Analytics} with the dynamical critical exponent. 
The results associated to the two short-ranged models are compatible, at late times, with $z=2$ as the system is diffusive. The long-range one-dimensional Ising model shows an intermediate-time anomalous superdiffusive behavior with $z=4/3$ when $\alpha = 1.5$, followed by a late-time diffusive behavior $z=2$.
When setting $\alpha = 1.1$, we instead get an anomalous value $z = 2 \alpha -1$ without late-time diffusion. 

\begin{figure}[t]
	\centering
    \includegraphics[width = 1\linewidth]{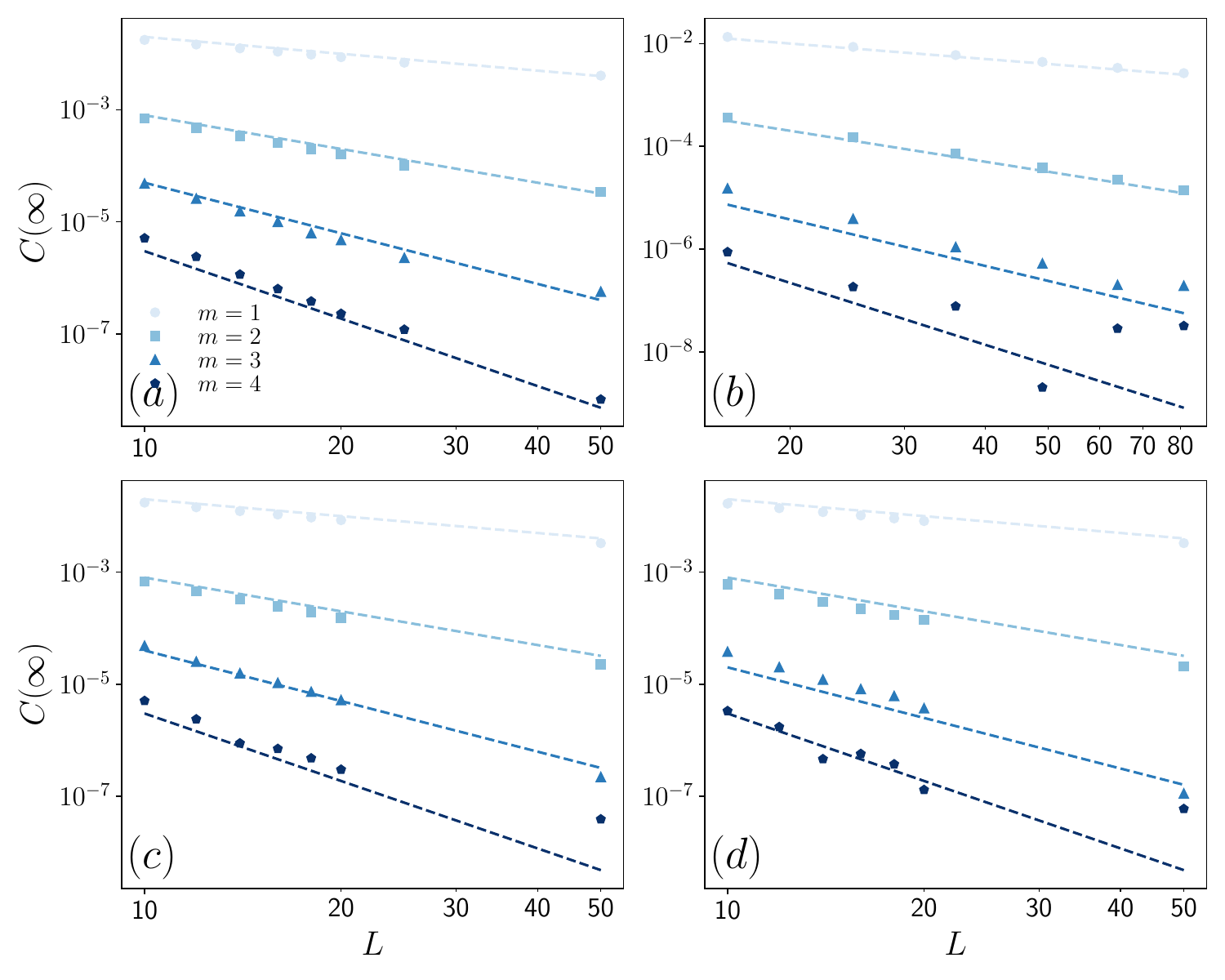}
	\caption{Long time average $C(\infty)$ versus $L$ for (a) one-dimensional mixed field Ising model;
    (b) two-dimensional transverse Ising model; (c) one-dimensional long range transverse Ising model $\alpha = 1.5$ and (d) one-dimensional long range transverse Ising model $\alpha = 1.1$. The dashed line indicates the analytical prediction $C(\infty)\propto L^{-m}$.} 
	\label{Fig:Ising:FinalSize}
\end{figure}

The second quantitative property is shown in Fig.~\ref{Fig:Ising:FinalSize} and shows that the infinite-time limit of the autocorrelation function has a scaling
\begin{equation}
 \underset{t\rightarrow\infty}{\lim}\langle \mathcal O(t) \mathcal O \rangle_c \sim 
 \frac{1}{L^m}
 \label{Eq:LateTimePlateau}
\end{equation}
and it is thus also determined by the thermodynamics of the model at zero energy density. The four different panels refer to the three models that we studied and highlight a rather universal scaling. The existence of a residual value in the late-time limit of autocorrelation functions that are expected to decay to zero in the thermodynamic limit has already been discussed in the context of chaotic quantum systems. The literature has highlighted the crucial difference between the algebraic decay to zero that we find also here in this classical framework and the exponential one, linking the two behaviors to the existence or absence of conservation laws~\cite{HuangBrandao_2019PRL, bp-22, bsrp-22}.

The two relations in Eqs.~\eqref{Eq:Ov:Rel} and~\eqref{Eq:LateTimePlateau}
are a remarkable regularity connecting thermodynamic properties and autocorrelations functions and the goal of the next section is to propose a theory that could explain them.

\section{Theoretical considerations}
\label{Sec:Analytics}

In this Section we propose a theoretical interpretation to the numerical results presented in Sec.~\ref{Sec:Numerics}, and summarized by Eqs.~\eqref{Eq:Ov:Rel} and~\eqref{Eq:LateTimePlateau}.

\subsection{Ergodicity}

We begin with some remarks on ergodicity for classical systems.
We have seen that in closed classical systems the Hamiltonian $H$ is always conserved and the thermal states are always stationary: 
we assume that no additional conserved quantities, independent from $H$, are present; to the best of our knowledge, this is the case for the three models studied in Sec.~\ref{Sec:Numerics}.

In order to formulate the notion of ergodicity,
let us consider
the observable $1_E$ that has support on the region of $\mathcal M$ with energy $E$ and that is defined by
\begin{equation}
\la 1_E\ra_{p} = \delta(E-H(p)), \quad p \in \mathcal{M}.
\end{equation}
We say that the system is \textit{ergodic} if any observable $\mathcal{O}$ spreads uniformly in time across the region of the phase space associated with the initial energy $E$. Mathematically, this means that the following relation holds at the observable level~\footnote{In Ref.~\cite{venuti2019ergodicity} this is called \textit{shell-ergodicity}, see Proposition~1 therein}:
\begin{equation}\label{eq:Obs_avg}
\int^T_0 \frac{dt}{T}
\mathcal{O}(t)
\;
\xrightarrow{T\to \infty}
\;
\int dE  \,
\frac{\langle 1_E \mathcal{O} \rangle_{\beta=0}}
{\langle 1_E \rangle_{\beta=0}}
\, \times 1_E.
\end{equation}
In particular, 
looking at Eq.~\eqref{eq:Obs_avg},
one identifies the microcanonical expectation value at energy $E$
\begin{equation}
 \mathcal{O}(E) :=
 \frac{\la 1_E \mathcal{O}\ra_{\beta=0}}{\la 1_E\ra_{\beta=0}}
 \label{Eq:O(E)}
\end{equation}
as the weight for the microcanonical distribution $1_E$. 

Eq.~\eqref{eq:Obs_avg} is crucial in our context, as it plays the role of the (diagonal part of the) ETH ansatz for quantum systems: physically, it means that the only parameter to describe stationary states in (closed) systems is the energy.
Whereas in Ref.~\cite{Capizzi_25}
we could prove the relaxation-overlap inequality in the quantum context assuming ETH, here we can prove it in the classical realm assuming ergodicity.
The analogy between the two notions, discussed in depth in Ref.~\cite{venuti2019ergodicity} but also in Refs.~\cite{mori_2017_ergodicity, 
Mori_Review_2018,
Alhambra_PRL2020}, is further analysed in Sec.~\ref{Sec:Discussion}.



An immediate consequence of ergodicity is the erasing of information after time evolution from a pure initial state:
\begin{equation}
\int^T_0 \frac{dt}{T}\la \mathcal{O}(t)\ra_p
\xrightarrow{T \rightarrow \infty}
\mathcal{O}(E = H(p)).
\label{Eq:Ergodicity:1}
\end{equation}
In summary, if one averages over large time windows, ergodicity ensures that the value of an observable always converges to that associated to the energy fixed by the initial conditions.
The expression in Eq.~\eqref{Eq:Ergodicity:1} can be regarded as a statement on the late-time dynamics of the observables, that converges to the microcanonical expectation value.

The writing employed in this section, where integrals over the phase space $\mathcal M$ are traded with integrals over energies, opens the windows to the introduction of a Boltzmann microcanonical entropy, defined as
\begin{equation}
 e^{S(E)} = \langle 1_E \rangle_{\beta=0},
\end{equation}
where the expectation value is taken over the flat measure defined in Sec.~\ref{Sec:II}.
Using this notation, little algebra allows us to express the late-time limit in Eq.~\eqref{Eq:Ergodicity:1} on a thermal state as:
\begin{equation}
\int_0^T \frac{dt}{T} \langle \mathcal O(t) \rangle_\beta \xrightarrow{T \to \infty}
\frac{\int dE \mathcal O(E)
e^{- \beta E + S(E)}}{\int dE e^{- \beta E + S(E)}}.
\label{Eq:Ergodicity:1bis}
\end{equation}
This relation demonstrates the convergence of any observable towards the stationary thermal expectation value.

The definitions and concepts above are general and do not require specific assumptions on the interactions of the underlying system. When we consider a many-body system, say locally interacting spins placed on a lattice of size $V$, the leading dependence on $V$ enters through the energy density $\varepsilon = E/V$ for local observables. Specifically, the microcanonical average $\mathcal{O}(\varepsilon=E/V)$ has a definite value in the limit $V\rightarrow \infty$; it coincides with that of a thermal ensemble at inverse temperature $\beta$ satisfying
\be\label{eq:beta_def}
\beta = \frac{dS}{dE},
\ee
expressing the local indistinguishability between canonical and microcanonical ensembles for infinitely large many-body systems. We warn the reader that finite-size effects, arising as algebraically small discrepancies between the two aforementioned ensembles, are present and are part of the discussion of the next two subsections. 

\subsection{Finite-size scaling in Eq.~\eqref{Eq:LateTimePlateau}}

In this, and in the next, subsections, we proceed with the proof of the relations in Eqs.~\eqref{Eq:Ov:Rel} and~\eqref{Eq:LateTimePlateau}.
Remarkably, assuming ergodicity it is possible to proceed with the proofs in the same way as we did in Ref.~\cite{Capizzi_25}, where we assumed ETH and worked in the quantum framework.
Hence, as anticipated, we will assume that our classical spin system is ergodic and that the energy is the only conserved quantity of the model.

We now focus on large, but finite, systems and study the autocorrelation function at finite temperature $\la \mathcal{O}(t)\mathcal{O}\ra_{\beta,c}$.
We take the late-time average, and, assuming ergodicity, we compute
\begin{equation}\label{eq:aut_avg}
\int^T_0 \frac{dt}{T}\la \mathcal{O}(t)\mathcal{O}\ra_{\beta,c}
\xrightarrow{T\rightarrow \infty}
\la \bar{\mathcal{O}^2}\ra_{\beta,c},
\end{equation}
where $\bar{\mathcal O}$ is defined by
\begin{equation}
\label{eq:O_avg}
\bar{\mathcal{O}} := \underset{T\rightarrow \infty}{\lim}\int^T_0\frac{dt}{T}\mathcal{O}(t).
\end{equation}
In order to obtain Eq.~\eqref{eq:aut_avg} we have only used the stationarity of the thermal state, together with the definition and the property $\mathcal{O}_1(t)\mathcal{O}_2(t) = (\mathcal{O}_1\mathcal{O}_2)(t)$, which is a direct consequence of Eq.~\eqref{eq:O_evol}. Specifically, this allows writing the argument of the integral in the l.h.s.~of Eq.~\eqref{eq:aut_avg} as 
$\langle \mathcal{O}(t+t')\mathcal{O}(t')\rangle_{\beta,c}$ for an arbitrary value of $t'$, but also as
an average over $t$ and $t'$:
\begin{equation}
\int^T_0 \frac{dt}{T}\la \mathcal{O}(t)\mathcal{O}\ra_{\beta,c}
= \int_0^T \int_0^T 
\frac{dt}{T} \frac{dt'}{T}
\langle \mathcal{O}(t+t')\mathcal{O}(t')\rangle_{\beta,c}.
\end{equation}
Eq.~\eqref{eq:aut_avg} follows after a simple change of variables.

Thanks to Eq.~\eqref{eq:aut_avg}, we now need to estimate the variance of the observable $\bar{\mathcal{O}}$, which is, in general, non-vanishing since the energy fluctuates in the thermal state.
We first assume ergodicity and, from Eqs.~\eqref{eq:Obs_avg} and~\eqref{Eq:Ergodicity:1bis}, we write explicitly
\begin{align}\label{eq-digeth}
\la \bar{\mathcal{O}^2}\ra_{\beta,c} =& \frac{\int dE e^{-\beta E+S(E)} \mathcal{O}(\varepsilon)^2}{\int dE e^{-\beta E+S(E)}} + \nonumber \\
&- \l\frac{\int dE e^{-\beta E+S(E)} \mathcal{O}(\varepsilon)}{\int dE e^{-\beta E+S(E)}}\r^2.
\end{align}
In the strict thermodynamic limit $V\rightarrow \infty$ the integral localizes at the saddle $\varepsilon = \varepsilon(\beta)$ satisfying Eq.~\eqref{eq:beta_def}, and the variance vanishes; to study the leading corrections around the saddle, we expand
\begin{equation}
\mathcal{O}(\varepsilon) \simeq \mathcal{O}(\varepsilon(\beta)) + \frac{1}{m!}\partial^m_\varepsilon\mathcal{O}(\varepsilon(\beta))(\varepsilon-\varepsilon(\beta))^m,
\label{Eq:Theory:Thermodynamics}
\end{equation}
where we have retained the first non-vanishing term in the Taylor expansion of $\mathcal O(\varepsilon)$ around $\varepsilon(\beta)$ (which is of order $m$).

Thanks to the central limit theorem for $H$, we estimate the variance of $(\varepsilon-\varepsilon(\beta))^m$ as
\be
\la (H/V-\varepsilon(\beta))^{2m}\ra_{\beta} - \l\la (H/V-\varepsilon(\beta))^{m}\ra_{\beta}\r^2 \sim \frac{1}{V^m},
\ee
using the argument that we employed in Ref.~\cite{Capizzi_25}. Putting everything together, we find that
\be
\underset{T\rightarrow \infty}{\lim}\int^T_0 \frac{dt}{T}\la \mathcal{O}(t)\mathcal{O}\ra_{\beta,c} \sim \frac{1}{V^m}.
\ee
This concludes the first proof, determining the link between the thermodynamic behaviour in Eq.~\eqref{Eq:Theory:Thermodynamics} and the scaling of the late-time plateau of the autocorrelation function with the system size in Eq.~\eqref{Eq:LateTimePlateau}.

In summary, the plateau of the autocorrelator at a given $\beta$ depends on the profile of $\mathcal{O}(\varepsilon)$ around $\varepsilon(\beta)$ and, in particular, on the leading term of the Taylor expansion. For practical purposes, the index $m$ can be extracted in a simpler way from the canonical expectation values as $\la \mathcal{O}\ra_{\beta'} - \la \mathcal{O}\ra_\beta\sim (\beta'-\beta)^m$. This is particularly useful when $\beta=0$, as explained in Ref. \cite{Capizzi_25}, since one identifies $m$ as the first integer such that $\la H^{m}\mathcal{O}\ra_{\beta=0}\neq 0$ (assuming $\la\mathcal{O}\ra_{\beta=0}=0$) and such a quantity can be efficiently computed; we will discuss the Ising model in Appendix~\ref{app:exp_val}.

\subsection{The relaxation-overlap inequality in Eq.~\eqref{Eq:Ov:Rel} and hydrodynamics}

In this section, we relate the finite-size plateau of the autocorrelator to the hydrodynamic tail. To do so, we report a physical argument that was first proposed in Ref.~\cite{Capizzi_25} for quantum many-body systems, since it applies straightforwardly for their corresponding classical counterpart.

First, the numerical studies of the autocorrelator $\la \mathcal{O}(t)\mathcal{O}\ra_c$ for local observables of many-body systems at finite-size $V$ presented in Sec.~\ref{Sec:Numerics} show that
for sufficiently small $t$
the thermodynamic limit is essentially achieved; it is the large-$t$ behaviour that suffers from finite-size effects. To understand the time scales where such effects become important, it is worth discussing the underlying hydrodynamics of the Hamiltonian, that is, in this context, the only conserved charge. For instance, given $h(\mathbf{x})$ the Hamiltonian density, this is expected to give rise~\cite{Spohn-12} to the following scaling law
\be\label{eq:h_2_point}
\la h(\mathbf{x},t)h(\mathbf{0},0)\ra_{\beta,c} \simeq \frac{1}{t^{1/z}}F\l \frac{|\mathbf{x}|}{t^{1/z}}\r;
\ee
here, $F$ is a universal function which depends on the transport properties, and $z$ is the dynamical critical exponent. In particular, the exponents appearing in Eq.~\eqref{eq:h_2_point} are matched so that once the expression in~\eqref{eq:h_2_point} is summed over $\mathbf x$ one gets a constant independent of time: this is a very important property that follows directly from energy conservation. 
For example, for one-dimensional short-range systems where the energy typically diffuses
\be
\partial_t h(x,t) \simeq D\partial^2_x h(x,t),
\ee
the dynamical exponent is $z=2$ and $F$ is a Gaussian. In general, from the hydrodynamics, one identifies a spatial region whose typical length scales as $\sim t^{1/z}$ where the energy density is spread at time $t$. As a consequence, before a timescale $t\lesssim V^{z/d}$ this region is smaller than the linear length of the setup and finite-size effects are expected to be irrelevant.

While the discussion above is restricted to charge densities, the same idea can be extended to a generic local observable $\mathcal{O}$. The key mechanism, which goes under the name of \textit{hydrodynamic projection}, is to project $\mathcal{O}$ over the charge densities (together with the powers and derivatives thereof, see Ref.~\cite{Doyon-22,Capizzi_25} for details): this projection can be characterized systematically in terms of the correlation functions of $h(\mathbf{x},t)$, while the rest, that is related to non-conserved fast decaying modes, is expected to give rise to subleading exponentially decaying contributions. Thus, the region where $\mathcal{O}$ spreads, relevant for the hydrodynamic tails of its autocorrelator, is approximately the same as the energy density. This implies that, for $t\lesssim V^{z/d}$ the autocorrelator decays as in the infinite volume limit, showing an asymptotic algebraic tail $t^{-\nu}$; at larger times it eventually saturates the plateau value, that has been estimated with $\sim V^{-m}$ in the previous sections.

Crucially, if one further assumes that the autocorrelator decays monotonically in time, which is a physically-motivated hypothesis for the data observed in Sec.~\ref{Sec:Numerics} provided one averages out the little transient oscillations, 
one can compare different times to get useful information. 
For instance, choosing a time of order $\sim V^{z/d}$ where the crossover occurs and the finite-size effects appear, one expects from monotonicity that $V^{-z\nu/d} \gtrsim V^{-m}$. 
The latter inequality implies $\nu \leq md/z$: this is precisely the relaxation-overlap inequality in Eq.~\eqref{Eq:Ov:Rel} that we presented in~\cite{Capizzi_25} for the quantum realm (see also Refs.~\cite{bbcp-21, bsrp-22, bp-22, DelacretazSciPost2020, wang2025eigenstate} for related results).

\section{Discussion: on the relation between ergodicity and ETH}
\label{Sec:Discussion}

We now elaborate on the relation between ergodicity and the diagonal part of ETH (see also the discussion in Ref.~\cite{venuti2019ergodicity} but also Refs.~\cite{mori_2017_ergodicity, Mori_Review_2018, Alhambra_PRL2020}). 
The point we want to stress here is that in generalizing our result from quantum to classical systems, it has been necessary to employ the standard notion of classical ergodicity whenever in the quantum setting we were employing the notion of ETH.
The analogy between these two concepts, which has been highlighted in previous works, is here found in a different context, namely as a tool for extending to the classical realm a result that was originally conceived in the quantum realm.

Let us begin by recalling the ETH: consider a quantum system with Hamiltonian $\hat H$ and an energy eigenstate $\ket{E_j}$, then the expectation value of a local observable $\hat{\mathcal O}$ is given by a smooth function that only depends on the energy density $\varepsilon = E_j/V$, namely
\begin{equation}
    \bra{E_j} \hat {\mathcal O} \ket{E_j} = \mathcal O(\varepsilon).
\end{equation}
In quantum many-body systems with a classical correspondence, the function $\mathcal O(\varepsilon)$ is expected to coincide with the classical one, given in Eq.~\eqref{Eq:EnergyDependence}; in the context of the spin systems considered here, it is reasonable to expect this to hold when comparing the classical model with a quantum model in the large-$S$ limit.

Let us now consider the quantum version of the expression given in Eq.~\eqref{eq:Obs_avg}:
\begin{equation}
    \int_0^T \frac{dt}{T} \hat{\mathcal O}(t) =
    \sum_{j,k} \int_0^T 
    \hspace{-0.25cm}
    e^{- i (E_j-E_k) t} \frac{dt}{T}
    \bra{E_j} \hat{\mathcal O} \ket{E_k} 
    \ket{E_j} \hspace{-0.1cm}\bra{E_k}.
\end{equation}
In the limit $T \to \infty$ all the phases average out and the expression becomes:
\begin{equation}\label{eq:diag}
   \lim_{T \to \infty} \int_0^T \frac{dt}{T}
   \hat{\mathcal O}(t) = \sum_j \bra{E_j} \hat{\mathcal O} \ket{E_j}
   \ket{E_j} \hspace{-0.1cm}\bra{E_j}.
\end{equation}
The analogy with the classical result in Eq.~\eqref{Eq:EnergyDependence} is apparent provided one identifies 
\begin{equation}
\sum_{j, E\in [E_0,E_0+\Delta E]} \ket{E_j}\bra{E_j} \leftrightarrow \int^{E_0+\Delta E}_{E_0} 1_E dE .
\end{equation}
Specifically, this holds whenever the sum (integral) is restricted to a given (not to small) energy shell, sufficiently large to host a large number of eigenstates making the classical/quantum correspondence meaningful. Using this fact, replacing the sum in \eqref{eq:diag} with partial sums over energy windows where $\bra{E_j} \hat {\mathcal O} \ket{E_j}$ is approximately constant, one gets the desired result, namely an expression that is fully analogous to Eq.~\eqref{eq:Obs_avg}.

As a last remark, we mention that ETH, besides the diagonal matrix elements, postulates the behaviour of the off-diagonal matrix elements as
\be
\bra{E_i}\hat{\mathcal{O}}\ket{E_j} \simeq e^{- \frac{S(\varepsilon)}{2}} 
   f_{\mathcal O}(\varepsilon, \omega) R_{ij}, \quad i\neq j,
\ee
with $\omega := E_i-E_j$, $S$ the microcanonical entropy and $R_{ij}$ a random matrix with zero mean and unit variance. The function $f_\mathcal{O}$ has an important role in the dynamics of the system since it is directly related to the finite-temperature autocorrelator as follows~\cite{dkpr-16,ms-19,sjhv-21}
\begin{align}
\label{eq:2pt_ETH1}
\la \hat{\mathcal{O}}(t)\hat{\mathcal{O}} \ra_{\beta,c} = \int d\omega e^{-\beta\omega/2+i\omega t} |f_\mathcal{O}\l \varepsilon(\beta),\omega\r|^2. 
\end{align}
In the classical framework, there is no notion analogous to the off-diagonal matrix elements; nonetheless, the autocorrelation function at equilibrium exists and has been thoroughly discussed throughout the text.
We conjecture that this might be the tool for finding an analogue of the off-diagonal ETH in the classical world.

\section{Conclusions and perspectives}
\label{Sec:Conclusions}

In this work, we studied the classical counterpart of the theory developed in Ref.~\cite{Capizzi_25}, where we linked the eigenstate thermalization hypothesis with hydrodynamics. 
The key point of this article is to replace the ETH of quantum systems by the notion of classical ergodicity. 
Such a correspondence allows us to infer a relaxation-overlap inequality for classical many-body systems. The theoretical predictions describe in a very satisfactory way the numerical results for large systems, corroborating with high precision the proposed scenario, arising from the hydrodynamics of the energy, common to both quantum and classical systems. 

Nonetheless, some questions regarding the link between ETH and ergodic many-body classical systems remain less clear. For example, it is not obvious what is the classical counterpart of the off-diagonal matrix elements with their random-matrix structure. Such a direction would be particularly interesting for extending this work to integrable systems, where the off-diagonal matrix elements are still under investigation \cite{Alba-15,ra-25,Essler-24}.

Another intriguing direction would be to understand the fate of our theory when ergodicity is weakly broken and additional many-body stationary states, besides the Gibbs ensembles, emerge. This scenario is the classical counterpart of quantum many-body scars, which are accompanied by violations of ETH. For instance, recent observations of interesting large-scale phenomena in quantum spin chains that go beyond conventional hydrodynamics \cite{Marche-25,mcf-25} present a challenge to incorporate into this work's framework.

Finally, we note that while this theory was developed to describe many-body systems at sufficiently high temperatures, distinct and relevant physics may emerge at lower temperatures. This is particularly pertinent for short-range spin chains in higher dimensions, where spontaneous symmetry breaking might occur, leading to the presence of dynamically disconnected phases. In such cases, one must re-evaluate the definition of ergodicity and consider the potential for long-range correlations. We hope to revisit this problem in future work.

\acknowledgments

We warmly thank X.~Xu for collaborating with us on a previous related article~\cite{Capizzi_25} and are grateful to L.~Foini, J.~Kurchan and S.~Pappalardi on the subject.
This work is supported by the ANR project LOQUST ANR-23-CE47-0006-02 (L.M. and L.C.). J.W.~acknowledges support from
the Deutsche
Forschungsgemeinschaft (DFG), under Grant No. 531128043, as well as under Grant
No.\ 397107022, No.\ 397067869, and No.\ 397082825 within the DFG Research
Unit FOR 2692, under Grant No.\ 355031190.
Additionally, we greatly acknowledge computing time on the HPC3 at the University of Osnabr\"{u}ck, granted by the DFG, under Grant No. 456666331.

\appendix

\section{Expectation values in the infinite temperature state}\label{app:exp_val}

In this appendix, we discuss the characterization of the expectation values of local observables in the infinite temperature state for classical spin systems. This is particularly relevant for the results of this work to compute the thermodynamical properties of local observables at small energy density $\varepsilon$. Specifically, we aim to compute the expectation values of monomials generated by the spin variables $\{(S^a_j)\}$. Since distinct sites are uncorrelated in the infinite temperature state, we focus on a given site and, from now on, we omit the site index $j$.

Before giving a quantitative characterization, it is worth providing a few symmetry considerations that ensure the vanishing of some quantities. For example, $\la S^z\ra=0$: this comes from the invariance of the Haar measure under the reflection $z\rightarrow -z$ and that $S^z\rightarrow -S^z$ under this transformation. Similarly, using reflection properties across the three axes, one can ensure that $\la (S^x)^{m_x} (S^y)^{m_y}(S^z)^{m_z}\ra$ vanishes whenever at least one of the three integers $m_a$ ($a=x,y,z$) is odd. 

As a consequence, the only non-vanishing monomials can only be those generated by $(S^a)^2$. These are explicitly non-vanishing: the reason is that $(S^a)^2$ is a positive observable, meaning that, as a function of the unit sphere, it is positive semidefinite and it only vanishes on a measure-zero subset; similarly, the product of positive observables is positive, as well as their expectation values. 

Finally, to give systematic predictions for the aforementioned expectation values, we can employ rotational invariance of the Haar measure. For example, using $(S^x)^2+(S^y)^2+(S^z)^2 =1$ and $\la(S^x)^2\ra = \la(S^y)^2\ra = \la(S^z)^2\ra$ (a consequence of the rotational symmetry), one easily gets $\la(S^a)^2\ra = 1/3$. In general, it is convenient to define the generating function $
\la e^{\boldsymbol{\mu} \cdot \mathbf{S}}\ra$ which gives directly the expectation values through its partial derivatives
\be
\begin{split}
\la (S^x)^{m_x}(S^y)^{m_y}(S^z)^{m_z}\ra = \partial^{m_x}_{\mu_x}\partial^{m_y}_{\mu_y}\partial^{m_z}_{\mu_z}\la e^{\boldsymbol{\mu}\cdot \mathbf{S}}\ra|_{\boldsymbol{\mu}=0}.
\end{split}
\ee
Using rotational invariance, we obtain $
\la e^{\boldsymbol{\mu}\cdot \mathbf{S}}\ra = \la e^{|\boldsymbol{\mu}| S^z}\ra$ and, from the expansion of the Haar measure in polar coordinates, we compute
\be
\begin{split}
\la e^{\boldsymbol{\mu}\cdot \mathbf{S}}\ra = \frac{\int^{\pi}_0 d\theta \sin \theta e^{|\boldsymbol{\mu}|\cos \theta}}{\int^{\pi}_0 d\theta \sin \theta} = \frac{\sinh|\boldsymbol{\mu}|}{|\boldsymbol{\mu}|}.
\end{split}
\ee
From that, the expectation values of monomials can be computed efficiently and, for example, we have
\be
\begin{split}
\la (S^x)^4\ra = 1/5, \quad \la (S^x)^6\ra = 1/7,\\
\la (S^x)^2(S^y)^2\ra = 1/15, \quad \la (S^x)^2(S^y)^4\ra = 1/35.
\end{split}
\ee

\section{Overlaps for the models considered in the article}\label{app:Overlaps}

In this appendix, we compute the overlaps between the local observables $\mathcal{O} =S_1^x$, $S_1^x S_{2}^x$, $S_1^x S_{2}^x S_{3}^x$, $S_1^x S_{2}^x S_{3}^x S_{4}^x$ and the Ising Hamiltonian in Eq.~\eqref{Eq:H:Ising} in the infinite temperature state $\beta=0$: specifically, we are interested in $\la H^{m'}\mathcal{O}\ra_{\beta=0}$ with a focus on the first values of $m'$ for which it vanishes. To do that, we expand $H^{m'}\mathcal{O}$ as a polynomial in $\{S^a_j\}$ and use the results of the Appendix~\ref{app:exp_val} for the infinite temperature state. 

First we consider $\mathcal{O} =S_1^x$, whose expectation value vanishes $\la \mathcal{O} \ra_{\beta=0}=0$. In the presence of a non-zero transverse field $h_x\neq 0$, a term $h_x(S^x_1)^2$ is generated in the expansion of $H\mathcal{O}$, which is the only one for which the expectation value does not vanish: as a consequence $\la H\mathcal{O}\ra_\beta \neq 0$, and the overlap order of $S_1^x$ is $m=1$.

More in general for the observable $\mathcal{O} = S^x_1\dots S^x_{m}$ we can get a non-zero expectation value of $H^{m'}\mathcal{O}$ whenever $H^{m'}$ contains terms that are odd (under change of sign) in $S^x_1,\dots,S^x_{m}$. This is possible when $m'=m$, since $H^m$ generates explicitly $h^m_xS^x_1\dots S^x_{m}$ in its polynomial expansion: on the other hand, one can easily check that it is not possible for $m'<m$. Therefore $\la H^m \mathcal{O}\ra_{\beta=0} \neq 0$  and the overlap order of $\mathcal{O}$ is precisely $m$.

This calculation can be easily extended to the other models considered in the article given in Eq.~\eqref{Eq:H:Ising:LR}
and 
Eq.~\eqref{Eq:H:Ising:2D}
obtaining the same result.

\bibliography{classical_Hydro}

\begin{thebibliography}{52}%
\makeatletter
\providecommand \@ifxundefined [1]{%
 \@ifx{#1\undefined}
}%
\providecommand \@ifnum [1]{%
 \ifnum #1\expandafter \@firstoftwo
 \else \expandafter \@secondoftwo
 \fi
}%
\providecommand \@ifx [1]{%
 \ifx #1\expandafter \@firstoftwo
 \else \expandafter \@secondoftwo
 \fi
}%
\providecommand \natexlab [1]{#1}%
\providecommand \enquote  [1]{``#1''}%
\providecommand \bibnamefont  [1]{#1}%
\providecommand \bibfnamefont [1]{#1}%
\providecommand \citenamefont [1]{#1}%
\providecommand \href@noop [0]{\@secondoftwo}%
\providecommand \href [0]{\begingroup \@sanitize@url \@href}%
\providecommand \@href[1]{\@@startlink{#1}\@@href}%
\providecommand \@@href[1]{\endgroup#1\@@endlink}%
\providecommand \@sanitize@url [0]{\catcode `\\12\catcode `\$12\catcode
  `\&12\catcode `\#12\catcode `\^12\catcode `\_12\catcode `\%12\relax}%
\providecommand \@@startlink[1]{}%
\providecommand \@@endlink[0]{}%
\providecommand \url  [0]{\begingroup\@sanitize@url \@url }%
\providecommand \@url [1]{\endgroup\@href {#1}{\urlprefix }}%
\providecommand \urlprefix  [0]{URL }%
\providecommand \Eprint [0]{\href }%
\providecommand \doibase [0]{https://doi.org/}%
\providecommand \selectlanguage [0]{\@gobble}%
\providecommand \bibinfo  [0]{\@secondoftwo}%
\providecommand \bibfield  [0]{\@secondoftwo}%
\providecommand \translation [1]{[#1]}%
\providecommand \BibitemOpen [0]{}%
\providecommand \bibitemStop [0]{}%
\providecommand \bibitemNoStop [0]{.\EOS\space}%
\providecommand \EOS [0]{\spacefactor3000\relax}%
\providecommand \BibitemShut  [1]{\csname bibitem#1\endcsname}%
\let\auto@bib@innerbib\@empty
\bibitem [{\citenamefont {Huang}(1987)}]{Huang1987}%
  \BibitemOpen
  \bibfield  {author} {\bibinfo {author} {\bibfnamefont {K.}~\bibnamefont
  {Huang}},\ }\href@noop {} {\emph {\bibinfo {title} {Statistical
  Mechanics}}},\ \bibinfo {edition} {2nd}\ ed.\ (\bibinfo  {publisher} {John
  Wiley \& Sons},\ \bibinfo {address} {New York},\ \bibinfo {year}
  {1987})\BibitemShut {NoStop}%
\bibitem [{\citenamefont {Simon}(1993)}]{Simon1993}%
  \BibitemOpen
  \bibfield  {author} {\bibinfo {author} {\bibfnamefont {B.}~\bibnamefont
  {Simon}},\ }\href@noop {} {\emph {\bibinfo {title} {The Statistical Mechanics
  of Lattice Gases}}}\ (\bibinfo  {publisher} {Princeton University Press},\
  \bibinfo {address} {Princeton},\ \bibinfo {year} {1993})\BibitemShut
  {NoStop}%
\bibitem [{\citenamefont {Pathria}(2016)}]{Pathria2016}%
  \BibitemOpen
  \bibfield  {author} {\bibinfo {author} {\bibfnamefont {R.}~\bibnamefont
  {Pathria}},\ }\href@noop {} {\emph {\bibinfo {title} {Statistical
  Mechanics}}}\ (\bibinfo  {publisher} {Butterworth-Heinemann},\ \bibinfo
  {year} {2016})\BibitemShut {NoStop}%
\bibitem [{\citenamefont {Khinchin}(1949)}]{Khinchin1949}%
  \BibitemOpen
  \bibfield  {author} {\bibinfo {author} {\bibfnamefont {A.~I.}\ \bibnamefont
  {Khinchin}},\ }\href@noop {} {\emph {\bibinfo {title} {Mathematical
  Foundations of Statistical Mechanics}}}\ (\bibinfo  {publisher} {Dover
  Publications},\ \bibinfo {address} {New York},\ \bibinfo {year}
  {1949})\BibitemShut {NoStop}%
\bibitem [{\citenamefont {Arnol'd}\ and\ \citenamefont
  {Avez}(1968)}]{ArnoldAvez1968}%
  \BibitemOpen
  \bibfield  {author} {\bibinfo {author} {\bibfnamefont {V.~I.}\ \bibnamefont
  {Arnol'd}}\ and\ \bibinfo {author} {\bibfnamefont {A.}~\bibnamefont {Avez}},\
  }\href@noop {} {\emph {\bibinfo {title} {Ergodic Problems of Classical
  Mechanics}}}\ (\bibinfo  {publisher} {W. A. Benjamin},\ \bibinfo {address}
  {New York},\ \bibinfo {year} {1968})\BibitemShut {NoStop}%
\bibitem [{\citenamefont {Ruelle}(1969)}]{Ruelle1969}%
  \BibitemOpen
  \bibfield  {author} {\bibinfo {author} {\bibfnamefont {D.}~\bibnamefont
  {Ruelle}},\ }\href@noop {} {\emph {\bibinfo {title} {Statistical Mechanics:
  Rigorous Results}}}\ (\bibinfo  {publisher} {W. A. Benjamin},\ \bibinfo
  {address} {New York},\ \bibinfo {year} {1969})\BibitemShut {NoStop}%
\bibitem [{\citenamefont {Petersen}(1983)}]{Petersen_1983}%
  \BibitemOpen
  \bibfield  {author} {\bibinfo {author} {\bibfnamefont {K.~E.}\ \bibnamefont
  {Petersen}},\ }\href@noop {} {\emph {\bibinfo {title} {Ergodic Theory}}},\
  Cambridge Studies in Advanced Mathematics\ (\bibinfo  {publisher} {Cambridge
  University Press},\ \bibinfo {year} {1983})\BibitemShut {NoStop}%
\bibitem [{\citenamefont {Ornstein}(1989)}]{Ornstein_1989}%
  \BibitemOpen
  \bibfield  {author} {\bibinfo {author} {\bibfnamefont {D.~S.}\ \bibnamefont
  {Ornstein}},\ }\href@noop {} {\emph {\bibinfo {title} {Ergodic theory,
  randomness, and" chaos"}}},\ Vol.\ \bibinfo {volume} {243}\ (\bibinfo
  {publisher} {American Association for the Advancement of Science},\ \bibinfo
  {year} {1989})\ pp.\ \bibinfo {pages} {182--187}\BibitemShut {NoStop}%
\bibitem [{\citenamefont {Kadanoff}\ and\ \citenamefont
  {Martin}(1963)}]{km-63}%
  \BibitemOpen
  \bibfield  {author} {\bibinfo {author} {\bibfnamefont {L.~P.}\ \bibnamefont
  {Kadanoff}}\ and\ \bibinfo {author} {\bibfnamefont {P.~C.}\ \bibnamefont
  {Martin}},\ }\bibfield  {title} {\bibinfo {title} {Hydrodynamic equations and
  correlation functions},\ }\href
  {https://doi.org/https://doi.org/10.1016/0003-4916(63)90078-2} {\bibfield
  {journal} {\bibinfo  {journal} {Annals of Physics}\ }\textbf {\bibinfo
  {volume} {24}},\ \bibinfo {pages} {419} (\bibinfo {year} {1963})}\BibitemShut
  {NoStop}%
\bibitem [{\citenamefont {El}\ and\ \citenamefont {Kamchatnov}(2005)}]{ek-05}%
  \BibitemOpen
  \bibfield  {author} {\bibinfo {author} {\bibfnamefont {G.~A.}\ \bibnamefont
  {El}}\ and\ \bibinfo {author} {\bibfnamefont {A.~M.}\ \bibnamefont
  {Kamchatnov}},\ }\bibfield  {title} {\bibinfo {title} {Kinetic equation for a
  dense soliton gas},\ }\href {https://doi.org/10.1103/PhysRevLett.95.204101}
  {\bibfield  {journal} {\bibinfo  {journal} {Phys. Rev. Lett.}\ }\textbf
  {\bibinfo {volume} {95}},\ \bibinfo {pages} {204101} (\bibinfo {year}
  {2005})}\BibitemShut {NoStop}%
\bibitem [{\citenamefont {Spohn}(2012)}]{Spohn-12}%
  \BibitemOpen
  \bibfield  {author} {\bibinfo {author} {\bibfnamefont {H.}~\bibnamefont
  {Spohn}},\ }\href@noop {} {\emph {\bibinfo {title} {Large scale dynamics of
  interacting particles}}}\ (\bibinfo  {publisher} {Springer Science \&
  Business Media},\ \bibinfo {year} {2012})\BibitemShut {NoStop}%
\bibitem [{\citenamefont {Belitz}\ \emph {et~al.}(2005)\citenamefont {Belitz},
  \citenamefont {Kirkpatrick},\ and\ \citenamefont {Vojta}}]{bkv-05}%
  \BibitemOpen
  \bibfield  {author} {\bibinfo {author} {\bibfnamefont {D.}~\bibnamefont
  {Belitz}}, \bibinfo {author} {\bibfnamefont {T.~R.}\ \bibnamefont
  {Kirkpatrick}},\ and\ \bibinfo {author} {\bibfnamefont {T.}~\bibnamefont
  {Vojta}},\ }\bibfield  {title} {\bibinfo {title} {How generic scale
  invariance influences quantum and classical phase transitions},\ }\href
  {https://doi.org/10.1103/RevModPhys.77.579} {\bibfield  {journal} {\bibinfo
  {journal} {Rev. Mod. Phys.}\ }\textbf {\bibinfo {volume} {77}},\ \bibinfo
  {pages} {579} (\bibinfo {year} {2005})}\BibitemShut {NoStop}%
\bibitem [{\citenamefont {Bettelheim}\ \emph {et~al.}(2006)\citenamefont
  {Bettelheim}, \citenamefont {Abanov},\ and\ \citenamefont
  {Wiegmann}}]{baw-06}%
  \BibitemOpen
  \bibfield  {author} {\bibinfo {author} {\bibfnamefont {E.}~\bibnamefont
  {Bettelheim}}, \bibinfo {author} {\bibfnamefont {A.~G.}\ \bibnamefont
  {Abanov}},\ and\ \bibinfo {author} {\bibfnamefont {P.}~\bibnamefont
  {Wiegmann}},\ }\bibfield  {title} {\bibinfo {title} {{Nonlinear Quantum Shock
  Waves in Fractional Quantum Hall Edge States}},\ }\href
  {https://doi.org/10.1103/PhysRevLett.97.246401} {\bibfield  {journal}
  {\bibinfo  {journal} {Phys. Rev. Lett.}\ }\textbf {\bibinfo {volume} {97}},\
  \bibinfo {pages} {246401} (\bibinfo {year} {2006})}\BibitemShut {NoStop}%
\bibitem [{\citenamefont {Nardin}\ and\ \citenamefont
  {Carusotto}(2023)}]{nc-23}%
  \BibitemOpen
  \bibfield  {author} {\bibinfo {author} {\bibfnamefont {A.}~\bibnamefont
  {Nardin}}\ and\ \bibinfo {author} {\bibfnamefont {I.}~\bibnamefont
  {Carusotto}},\ }\bibfield  {title} {\bibinfo {title} {{Linear and nonlinear
  edge dynamics of trapped fractional quantum Hall droplets}},\ }\href
  {https://doi.org/10.1103/PhysRevA.107.033320} {\bibfield  {journal} {\bibinfo
   {journal} {Phys. Rev. A}\ }\textbf {\bibinfo {volume} {107}},\ \bibinfo
  {pages} {033320} (\bibinfo {year} {2023})}\BibitemShut {NoStop}%
\bibitem [{\citenamefont {Lux}\ \emph {et~al.}(2014)\citenamefont {Lux},
  \citenamefont {M\"uller}, \citenamefont {Mitra},\ and\ \citenamefont
  {Rosch}}]{lmmr-14}%
  \BibitemOpen
  \bibfield  {author} {\bibinfo {author} {\bibfnamefont {J.}~\bibnamefont
  {Lux}}, \bibinfo {author} {\bibfnamefont {J.}~\bibnamefont {M\"uller}},
  \bibinfo {author} {\bibfnamefont {A.}~\bibnamefont {Mitra}},\ and\ \bibinfo
  {author} {\bibfnamefont {A.}~\bibnamefont {Rosch}},\ }\bibfield  {title}
  {\bibinfo {title} {Hydrodynamic long-time tails after a quantum quench},\
  }\href {https://doi.org/10.1103/PhysRevA.89.053608} {\bibfield  {journal}
  {\bibinfo  {journal} {Phys. Rev. A}\ }\textbf {\bibinfo {volume} {89}},\
  \bibinfo {pages} {053608} (\bibinfo {year} {2014})}\BibitemShut {NoStop}%
\bibitem [{\citenamefont {Moll}\ \emph {et~al.}(2016)\citenamefont {Moll},
  \citenamefont {Kushwaha}, \citenamefont {Nandi}, \citenamefont {Schmidt},\
  and\ \citenamefont {Mackenzie}}]{mknsm-16}%
  \BibitemOpen
  \bibfield  {author} {\bibinfo {author} {\bibfnamefont {P.~J.}\ \bibnamefont
  {Moll}}, \bibinfo {author} {\bibfnamefont {P.}~\bibnamefont {Kushwaha}},
  \bibinfo {author} {\bibfnamefont {N.}~\bibnamefont {Nandi}}, \bibinfo
  {author} {\bibfnamefont {B.}~\bibnamefont {Schmidt}},\ and\ \bibinfo {author}
  {\bibfnamefont {A.~P.}\ \bibnamefont {Mackenzie}},\ }\bibfield  {title}
  {\bibinfo {title} {{Evidence for hydrodynamic electron flow in PdCoO2}},\
  }\href {https://doi.org/https://doi.org/10.1126/science.aac8385} {\bibfield
  {journal} {\bibinfo  {journal} {Science}\ }\textbf {\bibinfo {volume}
  {351}},\ \bibinfo {pages} {1061} (\bibinfo {year} {2016})}\BibitemShut
  {NoStop}%
\bibitem [{\citenamefont {Crossno}\ \emph {et~al.}(2016)\citenamefont
  {Crossno}, \citenamefont {Shi}, \citenamefont {Wang}, \citenamefont {Liu},
  \citenamefont {Harzheim}, \citenamefont {Lucas}, \citenamefont {Sachdev},
  \citenamefont {Kim}, \citenamefont {Taniguchi}, \citenamefont {Watanabe}
  \emph {et~al.}}]{Crossno-16}%
  \BibitemOpen
  \bibfield  {author} {\bibinfo {author} {\bibfnamefont {J.}~\bibnamefont
  {Crossno}}, \bibinfo {author} {\bibfnamefont {J.~K.}\ \bibnamefont {Shi}},
  \bibinfo {author} {\bibfnamefont {K.}~\bibnamefont {Wang}}, \bibinfo {author}
  {\bibfnamefont {X.}~\bibnamefont {Liu}}, \bibinfo {author} {\bibfnamefont
  {A.}~\bibnamefont {Harzheim}}, \bibinfo {author} {\bibfnamefont
  {A.}~\bibnamefont {Lucas}}, \bibinfo {author} {\bibfnamefont
  {S.}~\bibnamefont {Sachdev}}, \bibinfo {author} {\bibfnamefont
  {P.}~\bibnamefont {Kim}}, \bibinfo {author} {\bibfnamefont {T.}~\bibnamefont
  {Taniguchi}}, \bibinfo {author} {\bibfnamefont {K.}~\bibnamefont {Watanabe}},
  \emph {et~al.},\ }\bibfield  {title} {\bibinfo {title} {{Observation of the
  Dirac fluid and the breakdown of the Wiedemann-Franz law in graphene}},\
  }\href {https://doi.org/10.1126/science.aad0343} {\bibfield  {journal}
  {\bibinfo  {journal} {Science}\ }\textbf {\bibinfo {volume} {351}},\ \bibinfo
  {pages} {1058} (\bibinfo {year} {2016})}\BibitemShut {NoStop}%
\bibitem [{\citenamefont {Lucas}\ and\ \citenamefont {Fong}(2018)}]{lf-18}%
  \BibitemOpen
  \bibfield  {author} {\bibinfo {author} {\bibfnamefont {A.}~\bibnamefont
  {Lucas}}\ and\ \bibinfo {author} {\bibfnamefont {K.~C.}\ \bibnamefont
  {Fong}},\ }\bibfield  {title} {\bibinfo {title} {Hydrodynamics of electrons
  in graphene},\ }\href {https://doi.org/10.1088/1361-648X/aaa274} {\bibfield
  {journal} {\bibinfo  {journal} {Journal of Physics: Condensed Matter}\
  }\textbf {\bibinfo {volume} {30}},\ \bibinfo {pages} {053001} (\bibinfo
  {year} {2018})}\BibitemShut {NoStop}%
\bibitem [{\citenamefont {Castro-Alvaredo}\ \emph {et~al.}(2016)\citenamefont
  {Castro-Alvaredo}, \citenamefont {Doyon},\ and\ \citenamefont
  {Yoshimura}}]{cdy-16}%
  \BibitemOpen
  \bibfield  {author} {\bibinfo {author} {\bibfnamefont {O.~A.}\ \bibnamefont
  {Castro-Alvaredo}}, \bibinfo {author} {\bibfnamefont {B.}~\bibnamefont
  {Doyon}},\ and\ \bibinfo {author} {\bibfnamefont {T.}~\bibnamefont
  {Yoshimura}},\ }\bibfield  {title} {\bibinfo {title} {Emergent hydrodynamics
  in integrable quantum systems out of equilibrium},\ }\href
  {https://doi.org/10.1103/PhysRevX.6.041065} {\bibfield  {journal} {\bibinfo
  {journal} {Phys. Rev. X}\ }\textbf {\bibinfo {volume} {6}},\ \bibinfo {pages}
  {041065} (\bibinfo {year} {2016})}\BibitemShut {NoStop}%
\bibitem [{\citenamefont {Bertini}\ \emph {et~al.}(2016)\citenamefont
  {Bertini}, \citenamefont {Collura}, \citenamefont {De~Nardis},\ and\
  \citenamefont {Fagotti}}]{bcdf-16}%
  \BibitemOpen
  \bibfield  {author} {\bibinfo {author} {\bibfnamefont {B.}~\bibnamefont
  {Bertini}}, \bibinfo {author} {\bibfnamefont {M.}~\bibnamefont {Collura}},
  \bibinfo {author} {\bibfnamefont {J.}~\bibnamefont {De~Nardis}},\ and\
  \bibinfo {author} {\bibfnamefont {M.}~\bibnamefont {Fagotti}},\ }\bibfield
  {title} {\bibinfo {title} {{Transport in out-of-equilibrium XXZ chains: Exact
  profiles of charges and currents}},\ }\href
  {https://doi.org/10.1103/PhysRevLett.117.207201} {\bibfield  {journal}
  {\bibinfo  {journal} {Phys. Rev. Lett.}\ }\textbf {\bibinfo {volume} {117}},\
  \bibinfo {pages} {207201} (\bibinfo {year} {2016})}\BibitemShut {NoStop}%
\bibitem [{\citenamefont {De~Nardis}\ \emph {et~al.}(2019)\citenamefont
  {De~Nardis}, \citenamefont {Bernard},\ and\ \citenamefont {Doyon}}]{dbd-19}%
  \BibitemOpen
  \bibfield  {author} {\bibinfo {author} {\bibfnamefont {J.}~\bibnamefont
  {De~Nardis}}, \bibinfo {author} {\bibfnamefont {D.}~\bibnamefont {Bernard}},\
  and\ \bibinfo {author} {\bibfnamefont {B.}~\bibnamefont {Doyon}},\ }\bibfield
   {title} {\bibinfo {title} {Diffusion in generalized hydrodynamics and
  quasiparticle scattering},\ }\href
  {https://doi.org/10.21468/SciPostPhys.6.4.049} {\bibfield  {journal}
  {\bibinfo  {journal} {SciPost Physics}\ }\textbf {\bibinfo {volume} {6}},\
  \bibinfo {pages} {049} (\bibinfo {year} {2019})}\BibitemShut {NoStop}%
\bibitem [{\citenamefont {Doyon}(2022{\natexlab{a}})}]{Doyon-22}%
  \BibitemOpen
  \bibfield  {author} {\bibinfo {author} {\bibfnamefont {B.}~\bibnamefont
  {Doyon}},\ }\bibfield  {title} {\bibinfo {title} {Diffusion and
  superdiffusion from hydrodynamic projections},\ }\href
  {https://doi.org/10.1007/s10955-021-02863-6} {\bibfield  {journal} {\bibinfo
  {journal} {Journal of Statistical Physics}\ }\textbf {\bibinfo {volume}
  {186}},\ \bibinfo {pages} {25} (\bibinfo {year}
  {2022}{\natexlab{a}})}\BibitemShut {NoStop}%
\bibitem [{\citenamefont {Doyon}(2022{\natexlab{b}})}]{Doyon-22a}%
  \BibitemOpen
  \bibfield  {author} {\bibinfo {author} {\bibfnamefont {B.}~\bibnamefont
  {Doyon}},\ }\bibfield  {title} {\bibinfo {title} {{Hydrodynamic projections
  and the emergence of linearised Euler equations in one-dimensional isolated
  systems}},\ }\href {https://doi.org/10.1007/s00220-022-04310-3} {\bibfield
  {journal} {\bibinfo  {journal} {Communications in Mathematical Physics}\
  }\textbf {\bibinfo {volume} {391}},\ \bibinfo {pages} {293} (\bibinfo {year}
  {2022}{\natexlab{b}})}\BibitemShut {NoStop}%
\bibitem [{\citenamefont {Capizzi}\ \emph {et~al.}(2025)\citenamefont
  {Capizzi}, \citenamefont {Wang}, \citenamefont {Xu}, \citenamefont {Mazza},\
  and\ \citenamefont {Poletti}}]{Capizzi_25}%
  \BibitemOpen
  \bibfield  {author} {\bibinfo {author} {\bibfnamefont {L.}~\bibnamefont
  {Capizzi}}, \bibinfo {author} {\bibfnamefont {J.}~\bibnamefont {Wang}},
  \bibinfo {author} {\bibfnamefont {X.}~\bibnamefont {Xu}}, \bibinfo {author}
  {\bibfnamefont {L.}~\bibnamefont {Mazza}},\ and\ \bibinfo {author}
  {\bibfnamefont {D.}~\bibnamefont {Poletti}},\ }\bibfield  {title} {\bibinfo
  {title} {Hydrodynamics and the eigenstate thermalization hypothesis},\ }\href
  {https://doi.org/10.1103/PhysRevX.15.011059} {\bibfield  {journal} {\bibinfo
  {journal} {Phys. Rev. X}\ }\textbf {\bibinfo {volume} {15}},\ \bibinfo
  {pages} {011059} (\bibinfo {year} {2025})}\BibitemShut {NoStop}%
\bibitem [{\citenamefont {Balachandran}\ \emph {et~al.}(2021)\citenamefont
  {Balachandran}, \citenamefont {Benenti}, \citenamefont {Casati},\ and\
  \citenamefont {Poletti}}]{bbcp-21}%
  \BibitemOpen
  \bibfield  {author} {\bibinfo {author} {\bibfnamefont {V.}~\bibnamefont
  {Balachandran}}, \bibinfo {author} {\bibfnamefont {G.}~\bibnamefont
  {Benenti}}, \bibinfo {author} {\bibfnamefont {G.}~\bibnamefont {Casati}},\
  and\ \bibinfo {author} {\bibfnamefont {D.}~\bibnamefont {Poletti}},\
  }\bibfield  {title} {\bibinfo {title} {{From the eigenstate thermalization
  hypothesis to algebraic relaxation of OTOCs in systems with conserved
  quantities}},\ }\href {https://doi.org/10.1103/PhysRevB.104.104306}
  {\bibfield  {journal} {\bibinfo  {journal} {Phys. Rev. B}\ }\textbf {\bibinfo
  {volume} {104}},\ \bibinfo {pages} {104306} (\bibinfo {year}
  {2021})}\BibitemShut {NoStop}%
\bibitem [{\citenamefont {Balachandran}\ \emph {et~al.}(2023)\citenamefont
  {Balachandran}, \citenamefont {Santos}, \citenamefont {Rigol},\ and\
  \citenamefont {Poletti}}]{bsrp-22}%
  \BibitemOpen
  \bibfield  {author} {\bibinfo {author} {\bibfnamefont {V.}~\bibnamefont
  {Balachandran}}, \bibinfo {author} {\bibfnamefont {L.~F.}\ \bibnamefont
  {Santos}}, \bibinfo {author} {\bibfnamefont {M.}~\bibnamefont {Rigol}},\ and\
  \bibinfo {author} {\bibfnamefont {D.}~\bibnamefont {Poletti}},\ }\bibfield
  {title} {\bibinfo {title} {{Slow relaxation of out-of-time-ordered
  correlators in interacting integrable and nonintegrable spin-$\frac{1}{2}$
  XYZ chains}},\ }\href {https://doi.org/10.1103/PhysRevB.107.235421}
  {\bibfield  {journal} {\bibinfo  {journal} {Phys. Rev. B}\ }\textbf {\bibinfo
  {volume} {107}},\ \bibinfo {pages} {235421} (\bibinfo {year}
  {2023})}\BibitemShut {NoStop}%
\bibitem [{\citenamefont {Balachandran}\ and\ \citenamefont
  {Poletti}(2022)}]{bp-22}%
  \BibitemOpen
  \bibfield  {author} {\bibinfo {author} {\bibfnamefont {V.}~\bibnamefont
  {Balachandran}}\ and\ \bibinfo {author} {\bibfnamefont {D.}~\bibnamefont
  {Poletti}},\ }\bibfield  {title} {\bibinfo {title} {{Relaxation exponents of
  OTOCs and overlap with local Hamiltonians}},\ }\href
  {https://doi.org/10.3390/e25010059} {\bibfield  {journal} {\bibinfo
  {journal} {Entropy}\ }\textbf {\bibinfo {volume} {25}},\ \bibinfo {pages}
  {59} (\bibinfo {year} {2022})}\BibitemShut {NoStop}%
\bibitem [{\citenamefont {Delacretaz}(2020)}]{DelacretazSciPost2020}%
  \BibitemOpen
  \bibfield  {author} {\bibinfo {author} {\bibfnamefont {L.~V.}\ \bibnamefont
  {Delacretaz}},\ }\bibfield  {title} {\bibinfo {title} {{Heavy operators and
  hydrodynamic tails}},\ }\href {https://doi.org/10.21468/SciPostPhys.9.3.034}
  {\bibfield  {journal} {\bibinfo  {journal} {SciPost Phys.}\ }\textbf
  {\bibinfo {volume} {9}},\ \bibinfo {pages} {034} (\bibinfo {year}
  {2020})}\BibitemShut {NoStop}%
\bibitem [{\citenamefont {Wang}\ \emph {et~al.}(2025)\citenamefont {Wang},
  \citenamefont {Mishra}, \citenamefont {Yang}, \citenamefont {Delacrétaz},\
  and\ \citenamefont {Pappalardi}}]{wang2025eigenstate}%
  \BibitemOpen
  \bibfield  {author} {\bibinfo {author} {\bibfnamefont {J.}~\bibnamefont
  {Wang}}, \bibinfo {author} {\bibfnamefont {R.}~\bibnamefont {Mishra}},
  \bibinfo {author} {\bibfnamefont {T.-H.}\ \bibnamefont {Yang}}, \bibinfo
  {author} {\bibfnamefont {L.~V.}\ \bibnamefont {Delacrétaz}},\ and\ \bibinfo
  {author} {\bibfnamefont {S.}~\bibnamefont {Pappalardi}},\ }\href
  {https://arxiv.org/abs/2505.06869} {\bibinfo {title} {Eigenstate
  thermalization hypothesis correlations via non-linear hydrodynamics}}
  (\bibinfo {year} {2025}),\ \Eprint {https://arxiv.org/abs/2505.06869}
  {arXiv:2505.06869 [cond-mat.stat-mech]} \BibitemShut {NoStop}%
\bibitem [{\citenamefont {Berry}(1977)}]{berry1977}%
  \BibitemOpen
  \bibfield  {author} {\bibinfo {author} {\bibfnamefont {M.~V.}\ \bibnamefont
  {Berry}},\ }\bibfield  {title} {\bibinfo {title} {Regular and irregular
  semiclassical wavefunctions},\ }\href
  {https://doi.org/10.1088/0305-4470/10/12/016} {\bibfield  {journal} {\bibinfo
   {journal} {Journal of Physics A: Mathematical and General}\ }\textbf
  {\bibinfo {volume} {10}},\ \bibinfo {pages} {2083} (\bibinfo {year}
  {1977})}\BibitemShut {NoStop}%
\bibitem [{\citenamefont {Deutsch}(1991)}]{Deutsch-91}%
  \BibitemOpen
  \bibfield  {author} {\bibinfo {author} {\bibfnamefont {J.~M.}\ \bibnamefont
  {Deutsch}},\ }\bibfield  {title} {\bibinfo {title} {Quantum statistical
  mechanics in a closed system},\ }\href
  {https://doi.org/10.1103/PhysRevA.43.2046} {\bibfield  {journal} {\bibinfo
  {journal} {Phys. Rev. A}\ }\textbf {\bibinfo {volume} {43}},\ \bibinfo
  {pages} {2046} (\bibinfo {year} {1991})}\BibitemShut {NoStop}%
\bibitem [{\citenamefont {Srednicki}(1999)}]{Srednicki-99}%
  \BibitemOpen
  \bibfield  {author} {\bibinfo {author} {\bibfnamefont {M.}~\bibnamefont
  {Srednicki}},\ }\bibfield  {title} {\bibinfo {title} {The approach to thermal
  equilibrium in quantized chaotic systems},\ }\href
  {https://iopscience.iop.org/article/10.1088/0305-4470/32/7/007} {\bibfield
  {journal} {\bibinfo  {journal} {Journal of Physics A: Mathematical and
  General}\ }\textbf {\bibinfo {volume} {32}},\ \bibinfo {pages} {1163}
  (\bibinfo {year} {1999})}\BibitemShut {NoStop}%
\bibitem [{\citenamefont {Venuti}\ and\ \citenamefont
  {Liu}(2019)}]{venuti2019ergodicity}%
  \BibitemOpen
  \bibfield  {author} {\bibinfo {author} {\bibfnamefont {L.~C.}\ \bibnamefont
  {Venuti}}\ and\ \bibinfo {author} {\bibfnamefont {L.}~\bibnamefont {Liu}},\
  }\href {https://arxiv.org/abs/1904.02336} {\bibinfo {title} {Ergodicity,
  eigenstate thermalization, and the foundations of statistical mechanics in
  quantum and classical systems}} (\bibinfo {year} {2019}),\ \Eprint
  {https://arxiv.org/abs/1904.02336} {arXiv:1904.02336 [cond-mat.stat-mech]}
  \BibitemShut {NoStop}%
\bibitem [{\citenamefont {Mori}(2017)}]{mori_2017_ergodicity}%
  \BibitemOpen
  \bibfield  {author} {\bibinfo {author} {\bibfnamefont {T.}~\bibnamefont
  {Mori}},\ }\bibfield  {title} {\bibinfo {title} {{Classical ergodicity and
  quantum eigenstate thermalization: Analysis in fully connected Ising
  ferromagnets}},\ }\href {https://doi.org/10.1103/PhysRevE.96.012134}
  {\bibfield  {journal} {\bibinfo  {journal} {Phys. Rev. E}\ }\textbf {\bibinfo
  {volume} {96}},\ \bibinfo {pages} {012134} (\bibinfo {year}
  {2017})}\BibitemShut {NoStop}%
\bibitem [{\citenamefont {Mori}\ \emph {et~al.}(2018)\citenamefont {Mori},
  \citenamefont {Ikeda}, \citenamefont {Kaminishi},\ and\ \citenamefont
  {Ueda}}]{Mori_Review_2018}%
  \BibitemOpen
  \bibfield  {author} {\bibinfo {author} {\bibfnamefont {T.}~\bibnamefont
  {Mori}}, \bibinfo {author} {\bibfnamefont {T.~N.}\ \bibnamefont {Ikeda}},
  \bibinfo {author} {\bibfnamefont {E.}~\bibnamefont {Kaminishi}},\ and\
  \bibinfo {author} {\bibfnamefont {M.}~\bibnamefont {Ueda}},\ }\bibfield
  {title} {\bibinfo {title} {Thermalization and prethermalization in isolated
  quantum systems: a theoretical overview},\ }\href
  {https://doi.org/10.1088/1361-6455/aabcdf} {\bibfield  {journal} {\bibinfo
  {journal} {Journal of Physics B: Atomic, Molecular and Optical Physics}\
  }\textbf {\bibinfo {volume} {51}},\ \bibinfo {pages} {112001} (\bibinfo
  {year} {2018})}\BibitemShut {NoStop}%
\bibitem [{\citenamefont {Alhambra}\ \emph {et~al.}(2020)\citenamefont
  {Alhambra}, \citenamefont {Riddell},\ and\ \citenamefont
  {Garc\'{\i}a-Pintos}}]{Alhambra_PRL2020}%
  \BibitemOpen
  \bibfield  {author} {\bibinfo {author} {\bibfnamefont {A.~M.}\ \bibnamefont
  {Alhambra}}, \bibinfo {author} {\bibfnamefont {J.}~\bibnamefont {Riddell}},\
  and\ \bibinfo {author} {\bibfnamefont {L.~P.}\ \bibnamefont
  {Garc\'{\i}a-Pintos}},\ }\bibfield  {title} {\bibinfo {title} {Time evolution
  of correlation functions in quantum many-body systems},\ }\href
  {https://doi.org/10.1103/PhysRevLett.124.110605} {\bibfield  {journal}
  {\bibinfo  {journal} {Phys. Rev. Lett.}\ }\textbf {\bibinfo {volume} {124}},\
  \bibinfo {pages} {110605} (\bibinfo {year} {2020})}\BibitemShut {NoStop}%
\bibitem [{\citenamefont {Schubert}\ \emph {et~al.}(2021)\citenamefont
  {Schubert}, \citenamefont {Richter}, \citenamefont {Jin}, \citenamefont
  {Michielsen}, \citenamefont {De~Raedt},\ and\ \citenamefont
  {Steinigeweg}}]{Schubert-21}%
  \BibitemOpen
  \bibfield  {author} {\bibinfo {author} {\bibfnamefont {D.}~\bibnamefont
  {Schubert}}, \bibinfo {author} {\bibfnamefont {J.}~\bibnamefont {Richter}},
  \bibinfo {author} {\bibfnamefont {F.}~\bibnamefont {Jin}}, \bibinfo {author}
  {\bibfnamefont {K.}~\bibnamefont {Michielsen}}, \bibinfo {author}
  {\bibfnamefont {H.}~\bibnamefont {De~Raedt}},\ and\ \bibinfo {author}
  {\bibfnamefont {R.}~\bibnamefont {Steinigeweg}},\ }\bibfield  {title}
  {\bibinfo {title} {Quantum versus classical dynamics in spin models: Chains,
  ladders, and square lattices},\ }\href
  {https://doi.org/10.1103/PhysRevB.104.054415} {\bibfield  {journal} {\bibinfo
   {journal} {Phys. Rev. B}\ }\textbf {\bibinfo {volume} {104}},\ \bibinfo
  {pages} {054415} (\bibinfo {year} {2021})}\BibitemShut {NoStop}%
\bibitem [{\citenamefont {Friedli}\ and\ \citenamefont
  {Velenik}(2017)}]{fv-17}%
  \BibitemOpen
  \bibfield  {author} {\bibinfo {author} {\bibfnamefont {S.}~\bibnamefont
  {Friedli}}\ and\ \bibinfo {author} {\bibfnamefont {Y.}~\bibnamefont
  {Velenik}},\ }\href {https://doi.org/10.1017/9781316882603} {\emph {\bibinfo
  {title} {Statistical Mechanics of Lattice Systems: A Concrete Mathematical
  Introduction}}}\ (\bibinfo  {publisher} {Cambridge University Press},\
  \bibinfo {year} {2017})\BibitemShut {NoStop}%
\bibitem [{\citenamefont {Nishikawa}\ and\ \citenamefont
  {Saito}(2025)}]{nishikawa2025energy-LR-Saito25}%
  \BibitemOpen
  \bibfield  {author} {\bibinfo {author} {\bibfnamefont {H.}~\bibnamefont
  {Nishikawa}}\ and\ \bibinfo {author} {\bibfnamefont {K.}~\bibnamefont
  {Saito}},\ }\bibfield  {title} {\bibinfo {title} {Energy diffusion in the
  long-range interacting spin systems},\ }\href
  {https://doi.org/10.48550/arXiv.2502.10139} {\bibfield  {journal} {\bibinfo
  {journal} {arXiv preprint arXiv:2502.10139}\ } (\bibinfo {year}
  {2025})}\BibitemShut {NoStop}%
\bibitem [{\citenamefont {Metropolis}\ \emph {et~al.}(1953)\citenamefont
  {Metropolis}, \citenamefont {Rosenbluth}, \citenamefont {Rosenbluth},
  \citenamefont {Teller},\ and\ \citenamefont
  {Teller}}]{metropolis1953equation}%
  \BibitemOpen
  \bibfield  {author} {\bibinfo {author} {\bibfnamefont {N.}~\bibnamefont
  {Metropolis}}, \bibinfo {author} {\bibfnamefont {A.~W.}\ \bibnamefont
  {Rosenbluth}}, \bibinfo {author} {\bibfnamefont {M.~N.}\ \bibnamefont
  {Rosenbluth}}, \bibinfo {author} {\bibfnamefont {A.~H.}\ \bibnamefont
  {Teller}},\ and\ \bibinfo {author} {\bibfnamefont {E.}~\bibnamefont
  {Teller}},\ }\bibfield  {title} {\bibinfo {title} {Equation of state
  calculations by fast computing machines},\ }\href
  {https://doi.org/10.1063/1.1699114} {\bibfield  {journal} {\bibinfo
  {journal} {The journal of chemical physics}\ }\textbf {\bibinfo {volume}
  {21}},\ \bibinfo {pages} {1087} (\bibinfo {year} {1953})}\BibitemShut
  {NoStop}%
\bibitem [{\citenamefont {Hastings}(1970)}]{hastings1970monte}%
  \BibitemOpen
  \bibfield  {author} {\bibinfo {author} {\bibfnamefont {W.~K.}\ \bibnamefont
  {Hastings}},\ }\bibfield  {title} {\bibinfo {title} {Monte carlo sampling
  methods using markov chains and their applications},\ }\href
  {http://www.jstor.org/stable/2334940} {\bibfield  {journal} {\bibinfo
  {journal} {Biometrika}\ }\textbf {\bibinfo {volume} {57}},\ \bibinfo {pages}
  {97} (\bibinfo {year} {1970})}\BibitemShut {NoStop}%
\bibitem [{\citenamefont {Yoshida}(1990)}]{Yoshida}%
  \BibitemOpen
  \bibfield  {author} {\bibinfo {author} {\bibfnamefont {H.}~\bibnamefont
  {Yoshida}},\ }\bibfield  {title} {\bibinfo {title} {Construction of higher
  order symplectic integrators},\ }\href
  {https://doi.org/https://doi.org/10.1016/0375-9601(90)90092-3} {\bibfield
  {journal} {\bibinfo  {journal} {Physics Letters A}\ }\textbf {\bibinfo
  {volume} {150}},\ \bibinfo {pages} {262} (\bibinfo {year}
  {1990})}\BibitemShut {NoStop}%
\bibitem [{\citenamefont {Huang}\ \emph {et~al.}(2019)\citenamefont {Huang},
  \citenamefont {Brand\~ao},\ and\ \citenamefont
  {Zhang}}]{HuangBrandao_2019PRL}%
  \BibitemOpen
  \bibfield  {author} {\bibinfo {author} {\bibfnamefont {Y.}~\bibnamefont
  {Huang}}, \bibinfo {author} {\bibfnamefont {F.~G. S.~L.}\ \bibnamefont
  {Brand\~ao}},\ and\ \bibinfo {author} {\bibfnamefont {Y.-L.}\ \bibnamefont
  {Zhang}},\ }\bibfield  {title} {\bibinfo {title} {Finite-size scaling of
  out-of-time-ordered correlators at late times},\ }\href
  {https://doi.org/10.1103/PhysRevLett.123.010601} {\bibfield  {journal}
  {\bibinfo  {journal} {Phys. Rev. Lett.}\ }\textbf {\bibinfo {volume} {123}},\
  \bibinfo {pages} {010601} (\bibinfo {year} {2019})}\BibitemShut {NoStop}%
\bibitem [{Note1()}]{Note1}%
  \BibitemOpen
  \bibinfo {note} {In Ref.~\cite {venuti2019ergodicity} this is called \protect
  \textit {shell-ergodicity}, see Proposition~1 therein}\BibitemShut {NoStop}%
\bibitem [{\citenamefont {D'Alessio}\ \emph {et~al.}(2016)\citenamefont
  {D'Alessio}, \citenamefont {Kafri}, \citenamefont {Polkovnikov},\ and\
  \citenamefont {Rigol}}]{dkpr-16}%
  \BibitemOpen
  \bibfield  {author} {\bibinfo {author} {\bibfnamefont {L.}~\bibnamefont
  {D'Alessio}}, \bibinfo {author} {\bibfnamefont {Y.}~\bibnamefont {Kafri}},
  \bibinfo {author} {\bibfnamefont {A.}~\bibnamefont {Polkovnikov}},\ and\
  \bibinfo {author} {\bibfnamefont {M.}~\bibnamefont {Rigol}},\ }\bibfield
  {title} {\bibinfo {title} {From quantum chaos and eigenstate thermalization
  to statistical mechanics and thermodynamics},\ }\href
  {https://doi.org/https://doi.org/10.1080/00018732.2016.1198134} {\bibfield
  {journal} {\bibinfo  {journal} {Advances in Physics}\ }\textbf {\bibinfo
  {volume} {65}},\ \bibinfo {pages} {239} (\bibinfo {year} {2016})}\BibitemShut
  {NoStop}%
\bibitem [{\citenamefont {Murthy}\ and\ \citenamefont
  {Srednicki}(2019)}]{ms-19}%
  \BibitemOpen
  \bibfield  {author} {\bibinfo {author} {\bibfnamefont {C.}~\bibnamefont
  {Murthy}}\ and\ \bibinfo {author} {\bibfnamefont {M.}~\bibnamefont
  {Srednicki}},\ }\bibfield  {title} {\bibinfo {title} {Bounds on chaos from
  the eigenstate thermalization hypothesis},\ }\href
  {https://doi.org/10.1103/PhysRevLett.123.230606} {\bibfield  {journal}
  {\bibinfo  {journal} {Phys. Rev. Lett.}\ }\textbf {\bibinfo {volume} {123}},\
  \bibinfo {pages} {230606} (\bibinfo {year} {2019})}\BibitemShut {NoStop}%
\bibitem [{\citenamefont {Sch{\"o}nle}\ \emph {et~al.}(2021)\citenamefont
  {Sch{\"o}nle}, \citenamefont {Jansen}, \citenamefont {Heidrich-Meisner},\
  and\ \citenamefont {Vidmar}}]{sjhv-21}%
  \BibitemOpen
  \bibfield  {author} {\bibinfo {author} {\bibfnamefont {C.}~\bibnamefont
  {Sch{\"o}nle}}, \bibinfo {author} {\bibfnamefont {D.}~\bibnamefont {Jansen}},
  \bibinfo {author} {\bibfnamefont {F.}~\bibnamefont {Heidrich-Meisner}},\ and\
  \bibinfo {author} {\bibfnamefont {L.}~\bibnamefont {Vidmar}},\ }\bibfield
  {title} {\bibinfo {title} {Eigenstate thermalization hypothesis through the
  lens of autocorrelation functions},\ }\href
  {https://doi.org/10.1103/PhysRevB.103.235137} {\bibfield  {journal} {\bibinfo
   {journal} {Phys. Rev. B}\ }\textbf {\bibinfo {volume} {103}},\ \bibinfo
  {pages} {235137} (\bibinfo {year} {2021})}\BibitemShut {NoStop}%
\bibitem [{\citenamefont {Alba}(2015)}]{Alba-15}%
  \BibitemOpen
  \bibfield  {author} {\bibinfo {author} {\bibfnamefont {V.}~\bibnamefont
  {Alba}},\ }\bibfield  {title} {\bibinfo {title} {Eigenstate thermalization
  hypothesis and integrability in quantum spin chains},\ }\href
  {https://doi.org/10.1103/PhysRevB.91.155123} {\bibfield  {journal} {\bibinfo
  {journal} {Phys. Rev. B}\ }\textbf {\bibinfo {volume} {91}},\ \bibinfo
  {pages} {155123} (\bibinfo {year} {2015})}\BibitemShut {NoStop}%
\bibitem [{\citenamefont {Rottoli}\ and\ \citenamefont {Alba}(2025)}]{ra-25}%
  \BibitemOpen
  \bibfield  {author} {\bibinfo {author} {\bibfnamefont {F.}~\bibnamefont
  {Rottoli}}\ and\ \bibinfo {author} {\bibfnamefont {V.}~\bibnamefont {Alba}},\
  }\href {https://arxiv.org/abs/2505.23602} {\bibinfo {title} {{Eigenstate
  Thermalization Hypothesis (ETH) for off-diagonal matrix elements in
  integrable spin chains}}} (\bibinfo {year} {2025}),\ \Eprint
  {https://arxiv.org/abs/2505.23602} {arXiv:2505.23602 [cond-mat.stat-mech]}
  \BibitemShut {NoStop}%
\bibitem [{\citenamefont {Essler}\ and\ \citenamefont
  {de~Klerk}(2024)}]{Essler-24}%
  \BibitemOpen
  \bibfield  {author} {\bibinfo {author} {\bibfnamefont {F.~H.~L.}\
  \bibnamefont {Essler}}\ and\ \bibinfo {author} {\bibfnamefont {A.~J. J.~M.}\
  \bibnamefont {de~Klerk}},\ }\bibfield  {title} {\bibinfo {title} {Statistics
  of matrix elements of local operators in integrable models},\ }\href
  {https://doi.org/10.1103/PhysRevX.14.031048} {\bibfield  {journal} {\bibinfo
  {journal} {Phys. Rev. X}\ }\textbf {\bibinfo {volume} {14}},\ \bibinfo
  {pages} {031048} (\bibinfo {year} {2024})}\BibitemShut {NoStop}%
\bibitem [{\citenamefont {March\'e}\ \emph {et~al.}(2025)\citenamefont
  {March\'e}, \citenamefont {Morettini}, \citenamefont {Mazza}, \citenamefont
  {Gotta},\ and\ \citenamefont {Capizzi}}]{Marche-25}%
  \BibitemOpen
  \bibfield  {author} {\bibinfo {author} {\bibfnamefont {A.}~\bibnamefont
  {March\'e}}, \bibinfo {author} {\bibfnamefont {G.}~\bibnamefont {Morettini}},
  \bibinfo {author} {\bibfnamefont {L.}~\bibnamefont {Mazza}}, \bibinfo
  {author} {\bibfnamefont {L.}~\bibnamefont {Gotta}},\ and\ \bibinfo {author}
  {\bibfnamefont {L.}~\bibnamefont {Capizzi}},\ }\bibfield  {title} {\bibinfo
  {title} {Exceptional stationary state in a dephasing many-body open quantum
  system},\ }\href {https://doi.org/10.1103/zn9v-k73w} {\bibfield  {journal}
  {\bibinfo  {journal} {Phys. Rev. Lett.}\ }\textbf {\bibinfo {volume} {135}},\
  \bibinfo {pages} {020406} (\bibinfo {year} {2025})}\BibitemShut {NoStop}%
\bibitem [{\citenamefont {Morettini}\ \emph {et~al.}(2025)\citenamefont
  {Morettini}, \citenamefont {Capizzi}, \citenamefont {Fagotti},\ and\
  \citenamefont {Mazza}}]{mcf-25}%
  \BibitemOpen
  \bibfield  {author} {\bibinfo {author} {\bibfnamefont {G.}~\bibnamefont
  {Morettini}}, \bibinfo {author} {\bibfnamefont {L.}~\bibnamefont {Capizzi}},
  \bibinfo {author} {\bibfnamefont {M.}~\bibnamefont {Fagotti}},\ and\ \bibinfo
  {author} {\bibfnamefont {L.}~\bibnamefont {Mazza}},\ }\href
  {https://arxiv.org/abs/2502.10387} {\bibinfo {title} {Unconventional
  transport in a system with a tower of quantum many-body scars}} (\bibinfo
  {year} {2025}),\ \Eprint {https://arxiv.org/abs/2502.10387} {arXiv:2502.10387
  [quant-ph]} \BibitemShut {NoStop}%
\end{thebibliography}%


\end{document}